\documentclass[12pt]{iopart}

\usepackage{graphicx}

\begin{document}

\title{Creation probabilities of hierarchical trees}{}
\author{A I Olemskoi$^1$, S S Borysov$^2$ and I A Shuda$^2$}

\address{$^1$Institute of Applied
Physics, Nat. Acad. Sci. of Ukraine, 58, Petropavlovskaya St.,
40030 Sumy, Ukraine}

\address{$^2$Sumy State University, 2, Rimskii-Korsakov St., 40007
Sumy, Ukraine}

\ead{alex@ufn.ru}

%\date{submitted to Journal of Physics A: }
\date{ \today }

\begin{abstract}
We consider both analytically and numerically creation conditions of diverse
hierarchical trees. A connection between the probabilities to create
hierarchical levels and the probability to associate these levels into united
structure is found. We argue a consistent probabilistic picture requires making
use of the deformed algebra. Our consideration is based on study of main types
of hierarchical trees, among which both regular and degenerate ones are studied
analytically, while the creation probabilities of the Fibonacci and free-scale
trees are determined numerically. We find a general expression for the creation
probability of an arbitrary tree and calculate the sum of terms of deformed
geometrical progression that appears at consideration of the degenerate tree.
\end{abstract}



\maketitle

\section{Introduction}\label{Sec.1}

The problem of the origin of hierarchy and its implications into physical,
biological, economical, ecological, social and other complex systems has a long
history which can be found in Refs. \cite{1a}, \cite{2a}, \cite{3a}, \cite{4a},
\cite{5a}, \cite{6a}, \cite{7a}, \cite{8a}, \cite{9a}, \cite{10a}, \cite{11a},
\cite{12a}, \cite{13a}. Along this line, one of the most striking
manifestations of hierarchy gives complex networks \cite{14a}. As is shown in
considerations of diverse systems, ranged from the World Wide Web \cite{15a} to
biological \cite{16a}, \cite{17a}, \cite{18a}, \cite{19a} and social
\cite{20a}, \cite{21a}, \cite{22a} networks, real networks are governed by
strict organizing principles displayed in the following properties: i) most
networks have a high degree of clustering; ii) many networks have been found to
be scale-free \cite{23a}, \cite{24a} that means the probability distribution
over node degrees, being the set of numbers of links with neighbors, follows
the power law. Moreover, many networks are modular: one can easily identify
groups of nodes that are highly interconnected with each other, but have only a
few or no links to nodes outside of the group to which they belong (in society
such modules represent groups of friends or coworkers \cite{25a}, in the WWW
denote communities with shared interests \cite{26a}, in the actor network they
characterize specific genres or simply individual movies). This clearly
identifiable modular organization is at the origin of the high clustering
coefficient seen in many real networks. In order to bring modularity, the high
degree of clustering and the scale-free topology under a single roof, we need
to assume that modules combine with each other in a hierarchical manner.

Formal basis of the theory of hierarchical structures is known by
the fact that hierarchically constrained objects are related to an
ultrametric space whose geometrical image is the Cayley tree with
nodes and branches corresponding to elementary cells and their
links \cite{Rammal}. One of the first theoretical pictures
\cite{Huberman} has been devoted to consideration of a diffusion
process on either uniformly or randomly multifurcating trees.
Consequent study of the hierarchical structures has shown
\cite{JETP_Let2000} their evolution is reduced to anomalous
diffusion process in ultrametric space that arrives at a
steady-state distribution over hierarchical levels, which
represents the Tsallis power law inherent in non-extensive systems
\cite{Tsallis}. A principle peculiarity of the Tsallis statistics
is known to be governed by a deformed algebra \cite{Borges}. Our
work is devoted to consideration of creation conditions of great
deal of variety of hierarchical trees on the basis of methods
developed initially at study of quantum groups \cite{QG}.

The outline of the paper is as follows. In Section \ref{Sec.2}, we
state a connection between probabilities to find hierarchical
levels with given set of effective energies and the probability to
associate these levels into united structure. We argue a
consistent probabilistic picture requires making use of the
deformed algebra, whose main rules are stated in Appendix A.
Further consideration is based on study of main types of
hierarchical trees depicted in Fig. \ref{fig1}:
\begin{figure}[!htb]
\centering
 \includegraphics[width=80mm,angle=0]{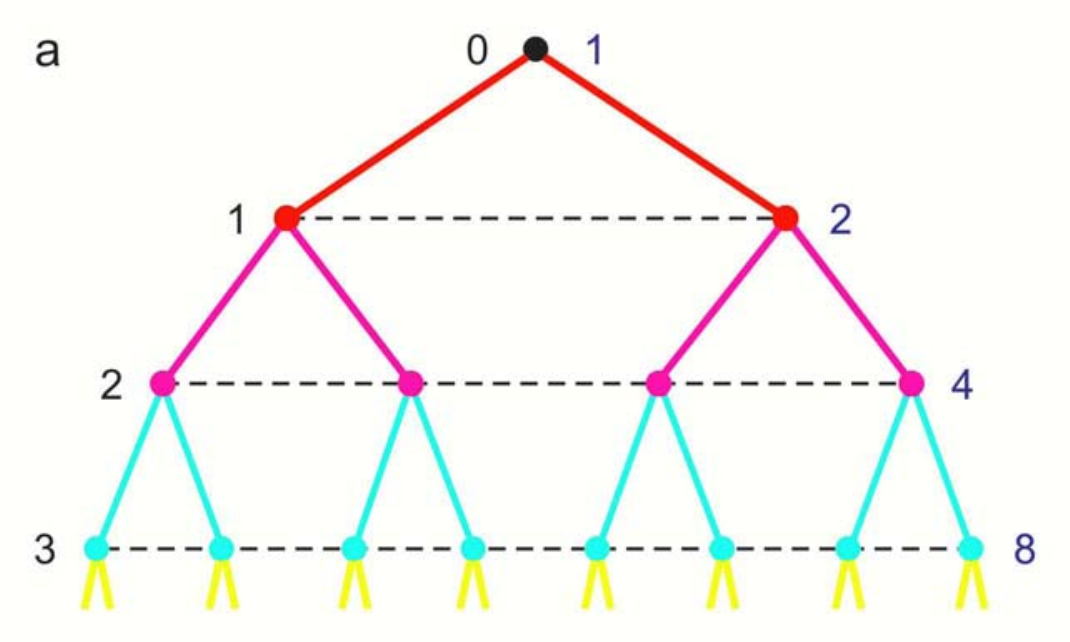}\\
 \includegraphics[width=80mm,angle=0]{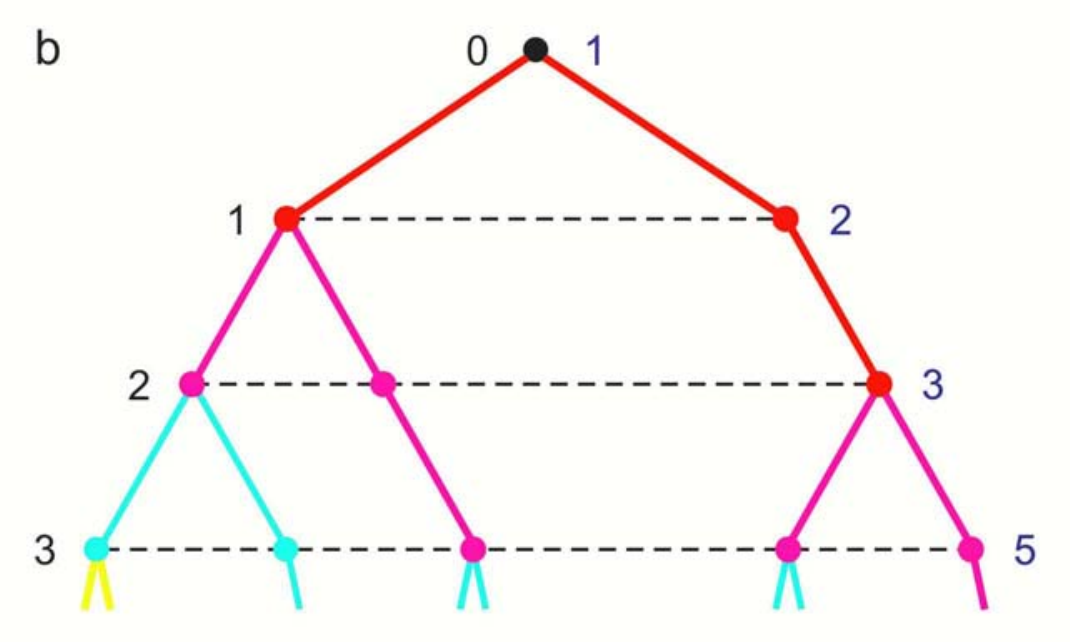}\\
 \includegraphics[width=80mm,angle=0]{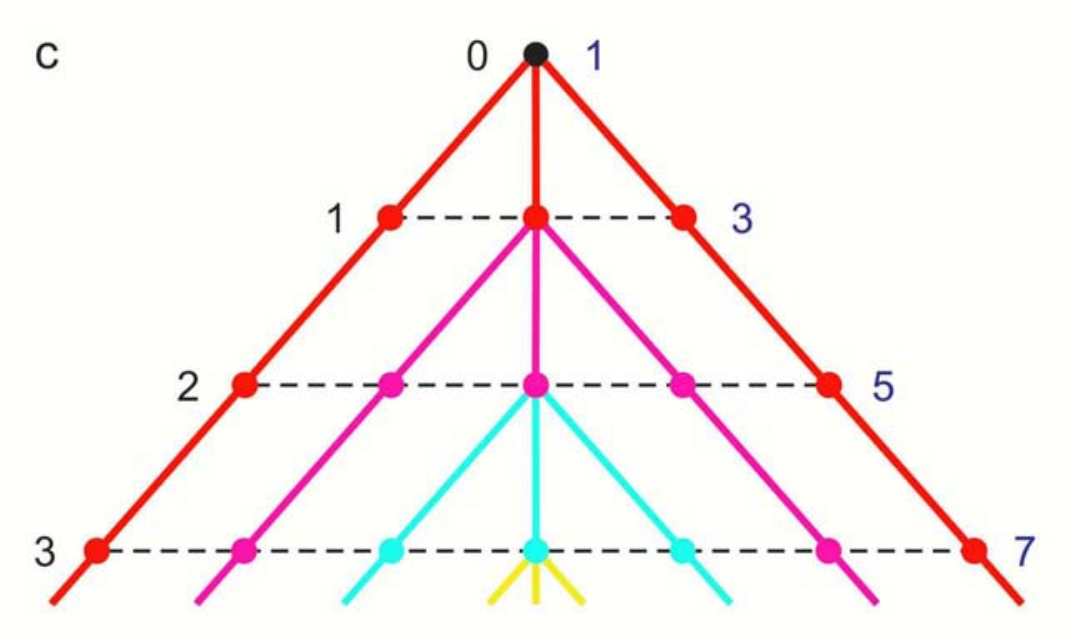}
  \caption{The main types of hierarchical trees (level numbers are indicated left,
node sum -- right). Top-down: regular tree with branching index
$b=2$, Fibonacci tree with $b=2$, and degenerate tree with $b=3$.}
 \label{fig1}
\end{figure}
Sections \ref{Sec.3} and \ref{Sec.4} are devoted to analytical definition of
the creation probabilities of both regular and degenerate trees, while in
Section \ref{Sec.5} we find these for both Fibonacci and free-scale trees
numerically. The case of an arbitrary tree is considered in Section \ref{Sec.6}
and Section \ref{Sec.7} is devoted to discussion of obtained results. Appendix
B contains details of calculations of the sum of terms of deformed geometrical
progression that appears at consideration of degenerate tree.

\section{Defining creation probability of hierarchical structure}\label{Sec.2}

Let us consider a hierarchical structure comprising of $n>1$
levels $l=0,1,\dots,n$ characterized by energy barrier heights
$\epsilon_l$ and total height $\epsilon_n$ connected with the
natural additivity assumption
\begin{equation}
\epsilon_n:=\sum_{l=0}^{n}\epsilon_l.
 \label{2}
\end{equation}
The principle peculiarity of hierarchical ensembles is known to be presented
with the Tsallis' thermostatistics \cite{Tsallis} where the $l$th hierarchical
level is related to the probability \cite{JETP_Let2000}
\begin{equation}
p_l=\exp_q\left(-\frac{\epsilon_l}{\Delta}\right)
 \label{1}
\end{equation}
characterized by the deformed exponential (\ref{6}) with
dispersion $\Delta$ and height of energy barrier $\epsilon_l$.
Self-consistent probabilistic picture of hierarchical ensembles is
reached if one proposes that, in contrast to the additivity rule
(\ref{2}), the normalization condition
\begin{equation}
p_0\oplus_q p_1\oplus_q\dots\oplus_q p_n=1
 \label{1a}
\end{equation}
is deformed to fix the top level probability $p_0$ according to the summation
rule (\ref{77}).

Along this line, one should set the probability $P_n$ related to
the $n$-level hierarchical structure determines the total height
of energy barrier $\epsilon_n=-\Delta\ln_q(P_n)$ through the
deformed logarithm (\ref{6}). Then, the condition (\ref{2})
arrives at the additivity of these logarithms:
\begin{equation}
\ln_q(P_n)=\sum\limits_{l=0}^{n}\ln_q(p_l).
 \label{4}
\end{equation}
In accordance with the first rule (\ref{8}), this equation means the
probability relation
\begin{equation}
P_n:=p_0\otimes_q p_1\otimes_q p_2\otimes_q\dots\otimes_q p_n.
 \label{5}
\end{equation}
Thus, in contrast to ordinary statistical systems, the creation probability
$P_n$ of a hierarchical structure equals to the {\it deformed} production of
specific probabilities $p_l$ related to levels $l=0,1,\dots,n$. As the
production definition (\ref{7}) shows, growth of the deformation parameter
$q>1$ increases essentially the probability (\ref{5}) in comparison with the
usual value at $q=1$. From physical point of view, above deformation of the
factorization rule for independent probabilities recovers the additivity
condition (\ref{2}) for corresponding heights of the energy barriers within the
Tsallis' thermostatistics.

With accounting (\ref{6}), Eq. (\ref{4}) arrives at the explicit form of the
creation probability of a hierarchical structure:
\begin{equation}
P_n=\exp_q\left[\frac{\sum_{l=0}^{n}p_l^{1-q}-(n+1)}{1-q}\right]
=\left(\sum_{l=0}^{n}p_l^{1-q}-n\right)_+^{\frac{1}{1-q}}.
 \label{11}
\end{equation}
Here, the last expression follows directly from the deformed production
(\ref{5}) with accounting the rule (\ref{7}). The relations (\ref{11}) mean the
decrease of the creation probability with growing hierarchical tree in
accordance with the difference equation
\begin{equation}
P_{n-1}^{1-q}-P_n^{1-q}=1-p_n^{1-q}.
 \label{13}
\end{equation}
In non-deformed limit $q\to 1$, relations (\ref{5}) and (\ref{11}) are reduced
to the ordinary rule $P_n=\prod_{l=0}^n p_l$ (respectively, Eq. (\ref{13})
reads $P_{n}/P_{n-1}=p_n$), while at $q=2$ the creation probability (\ref{11})
takes a maximal value.

A principle peculiarity of the above scheme is that level energies
$\epsilon_l$ remain to be additive values because creation of
hierarchical structure does not break the law of the energy
conservation. However, the hierarchy deforms essentially the
probability relations (\ref{1a}), (\ref{5}), (\ref{11}) and
(\ref{13}) due to appearance of coupling between level
probabilities $p_l$.

According to Eq. (\ref{11}) the consequent step in definition of
the creation probability $P_n$ of a hierarchical structure comes
to determination of set of probabilities $\{p_l\}_0^n$ related to
different hierarchical levels. Let us consider first the simplest
case of the regular tree depicted in Fig. \ref{fig1}(a).

\section{Regular tree}\label{Sec.3}

Let us consider a regular tree whose nodes multifurcate on certain level $l$
with constant branching index $b>1$ to generate a set of the $N_l=b^l$ nodes
determined with inherent probabilities $\pi=p_0/N_l=p_0b^{-l}$ where $p_0$ is
their top magnitude being normalization constant. Within naive proposition, one
could permit additivity of the node probabilities to arrive at the total
probability of the $l$ level realization to be $p_l:=N_l\pi=p_0$. Thus, within
the condition of additivity of the node probabilities, related values
$p_l=p_0=(n+1)^{-1}$ for all levels appear to be non-dependent of their numbers
$l=0,1,\dots,n$.

To escape such trivial situation we propose to replace above additive
connection of the level probability $p_l$ with the node value $\pi$ by the
following deformed equalities:
\begin{equation}\label{a}
\frac{p_l}{p_0}:=\underbrace{\frac{\pi}{p_0}\oplus_q\frac{\pi}{p_0}\oplus_q
\dots\oplus_q\frac{\pi}{p_0}}_{N_l}\equiv
N_l\odot_q\frac{\pi}{p_0}=b^{l}\odot_q b^{-l}.
\end{equation}
Presenting here the deformed sum of $N_l$ identical terms with help of the
formula (\ref{77a}), one obtains the level distribution required in the
binomial form
\begin{equation}\label{aa}
p_l=p_0\frac{[1+(1-q)b^{-l}]_+^{b^l}-1}{1-q}.
\end{equation}
In the limit $q\to 1$, it is simplified into expression
\begin{equation}\label{b}
p_l\simeq p_0\left[1+\frac{1-q}{2}\big(1-b^{-l}\big)\right]
\end{equation}
that shows exponentially fast variation with growth of the level number $l\geq
1$. In the cases $l\gg 1$ or $b\gg 1$, the probability (\ref{aa})
reaches the limit value
\begin{equation}
p_\infty=\frac{{\rm e}^{1-q}-1}{1-q}p_0=p_0\ln_q{\rm e}
 \label{c}
\end{equation}
being $p_\infty>p_0$ at deformation $q<1$ and $p_\infty<p_0$ at
deformation $q>1$. Inserting Eq. (\ref{c}) into Eq. (\ref{11})
leads to the following expression for the creation probability of
a regular tree:
\begin{equation}
P_n=\left[p_0^{1-q}\left(1+n\left(\ln_q{\rm
e}\right)^{1-q}\right)-n\right]^{\frac{1}{1-q}}.
 \label{Pr}
\end{equation}
Respectively, the deformed normalization condition related to the limit $b\gg
1$ takes the form
\begin{eqnarray}  \label{Pr}
%\begin{split}
p_0\oplus_q\left[n\odot_q\left(p_0\ln_q{\rm e}\right)\right]\nonumber\\
=\frac{\left[1+p_0\left(1-q\right)\right]\left[1+p_0\left(1-q\right)\ln_q{\rm
e}\right]^n-1}{1-q}=1.
%\end{split}
\end{eqnarray}

\begin{figure}[!htb]
\centering
 \includegraphics[width=85mm,angle=0]{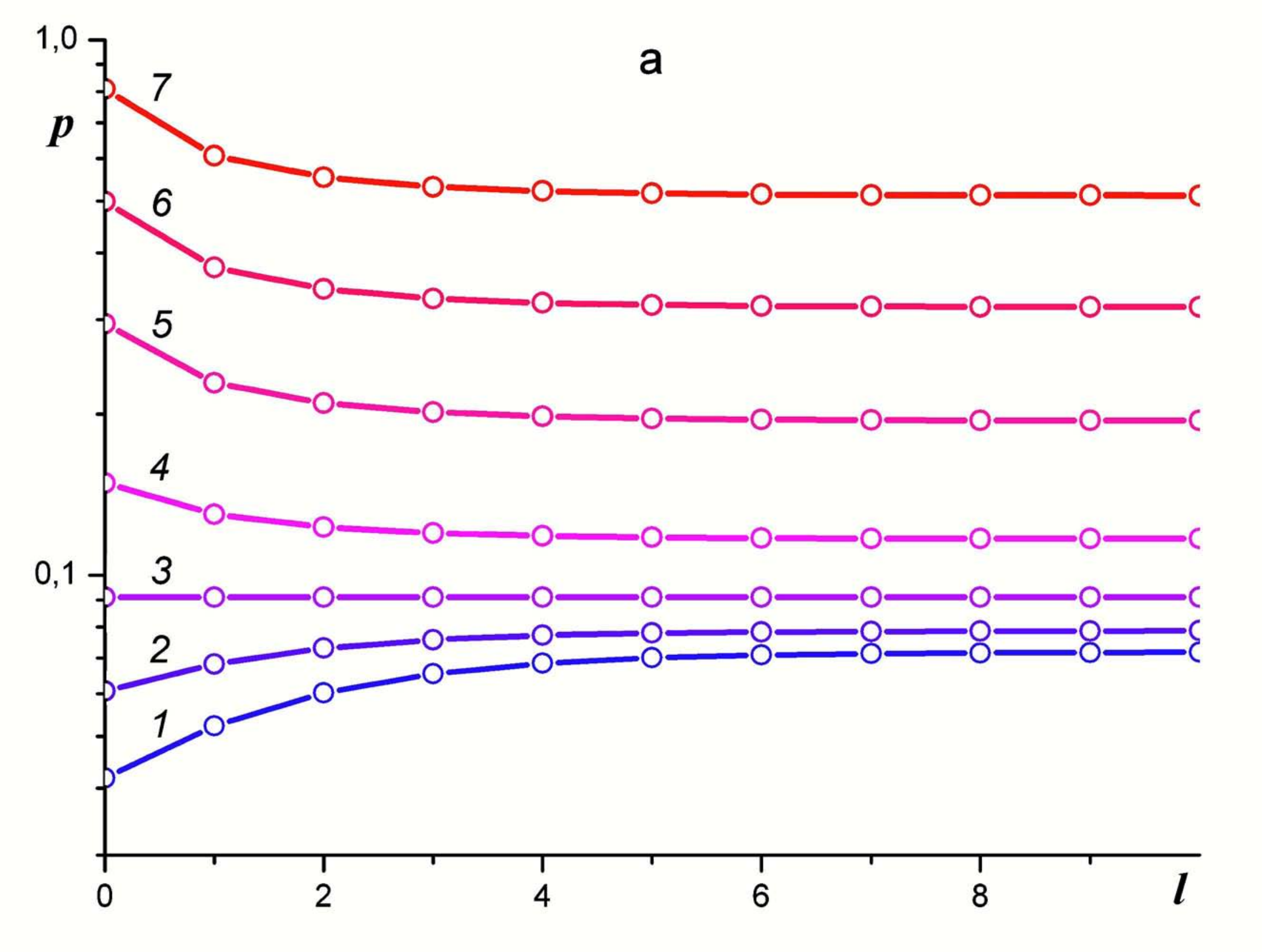}\\
 \includegraphics[width=85mm,angle=0]{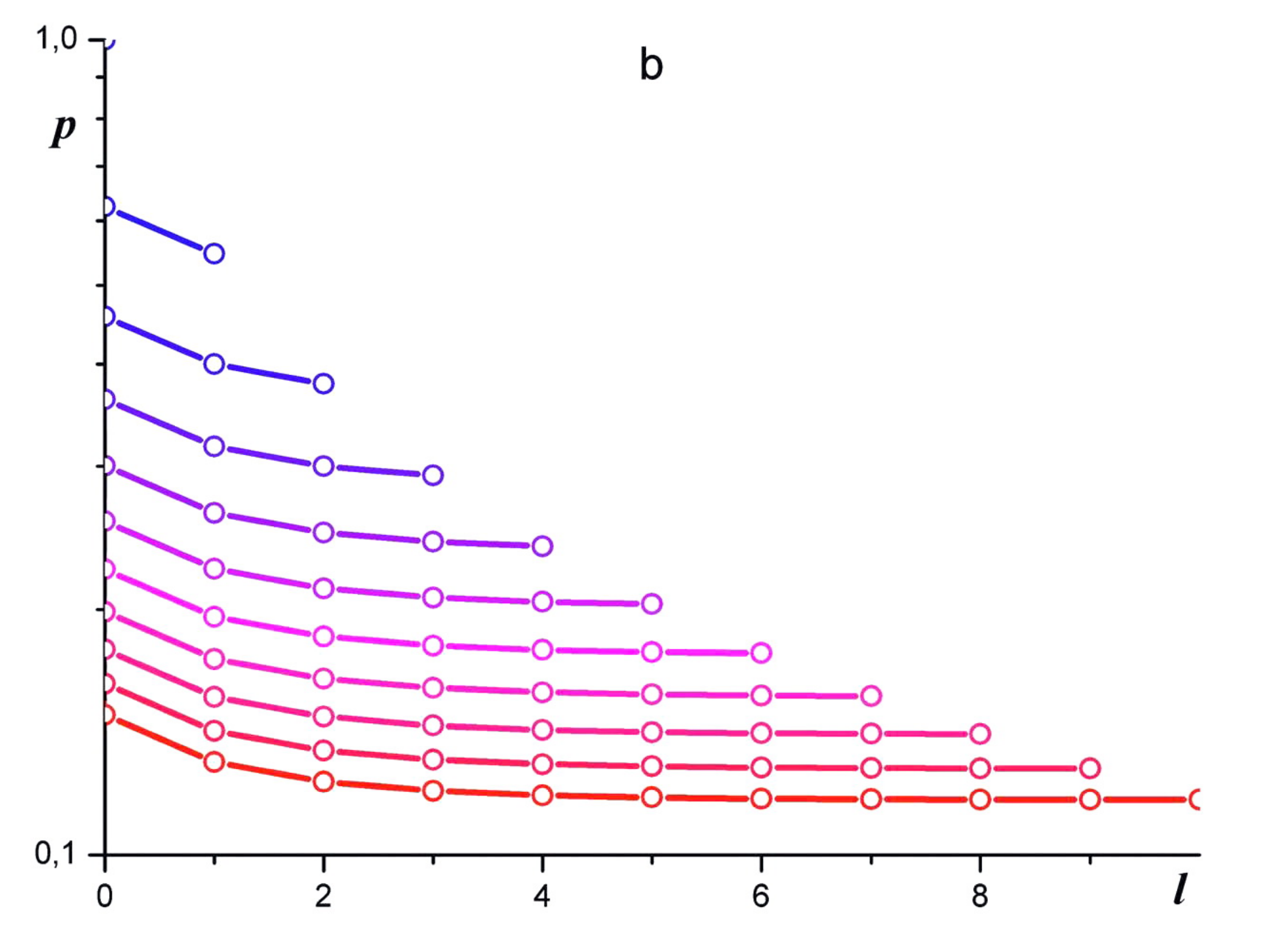}\\
 \includegraphics[width=85mm,angle=0]{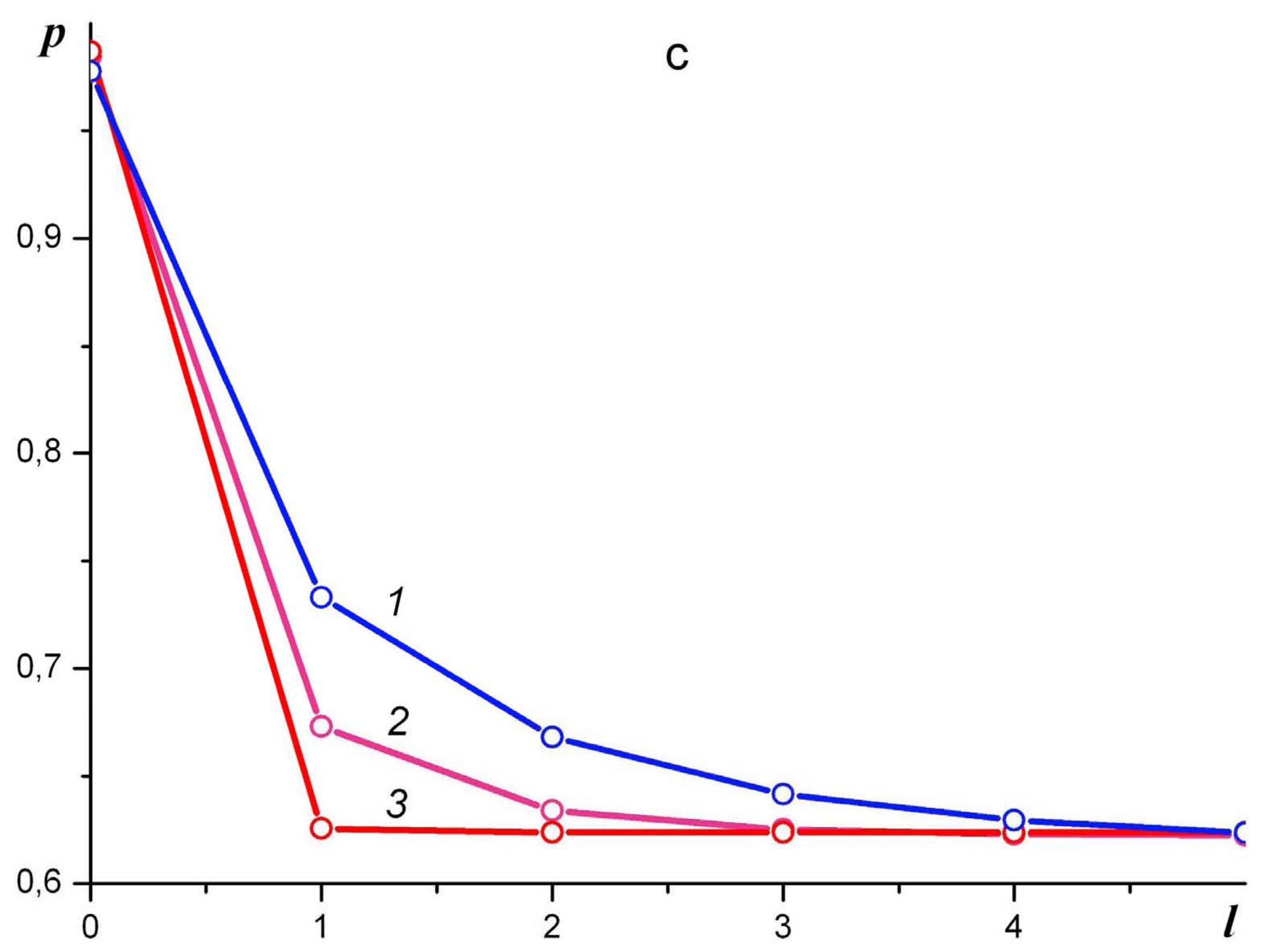}\\
 \caption{Probability distribution over hierarchical levels
 of the regular tree as function of the
level number at: (a) $b=2$, $n=10$ and
$q=10^{-4},0.5,1.0,1.5,1.9,1.99,1.9999$ (curves 1-7,
respectively); (b) $b=2$, $q=1.5$ and $n=1,2,\dots,10$ (curves
top-down, respectively); (c) $q=1.9999$, $n=5$ and $b=2,4,100$
(curves 1-3, respectively).} \label{fig2}
\end{figure}

In accordance with above consideration, Fig. \ref{fig2}(a) shows
the probability
(\ref{aa}) increases with growing number $l$ of hierarchical level at $q<1$ and
decays at $q>1$. From physical point of view, the creation probability of a
deeper hierarchical level should be less than this for upper levels, so that
one ought to conclude that the case $q>1$ is meaningful only.

In this case, with growing total number of hierarchical levels
$n$, the probability distribution (\ref{aa}) normalized with the
condition (\ref{1a}) decays as it is shown in Fig. \ref{fig2}(b).
Characteristically, the form of this distribution depends very
slightly on both deformation parameter $q$ and branching index $b$
excluding the domain $2-q\ll1$. According to Fig. \ref{fig2}(c),
within this domain, the probability distribution over hierarchical
levels decays not so sharply at small values of the branching
index $b$. With large growing the parameter $b\gg1$, the
dependence $p_l$ decreases more sharply to reach exponentially
fast the minimum value (\ref{c}) that is independent of the
branching index $b$.

\begin{figure}[!htb]
\centering
 \includegraphics[width=80mm,angle=0]{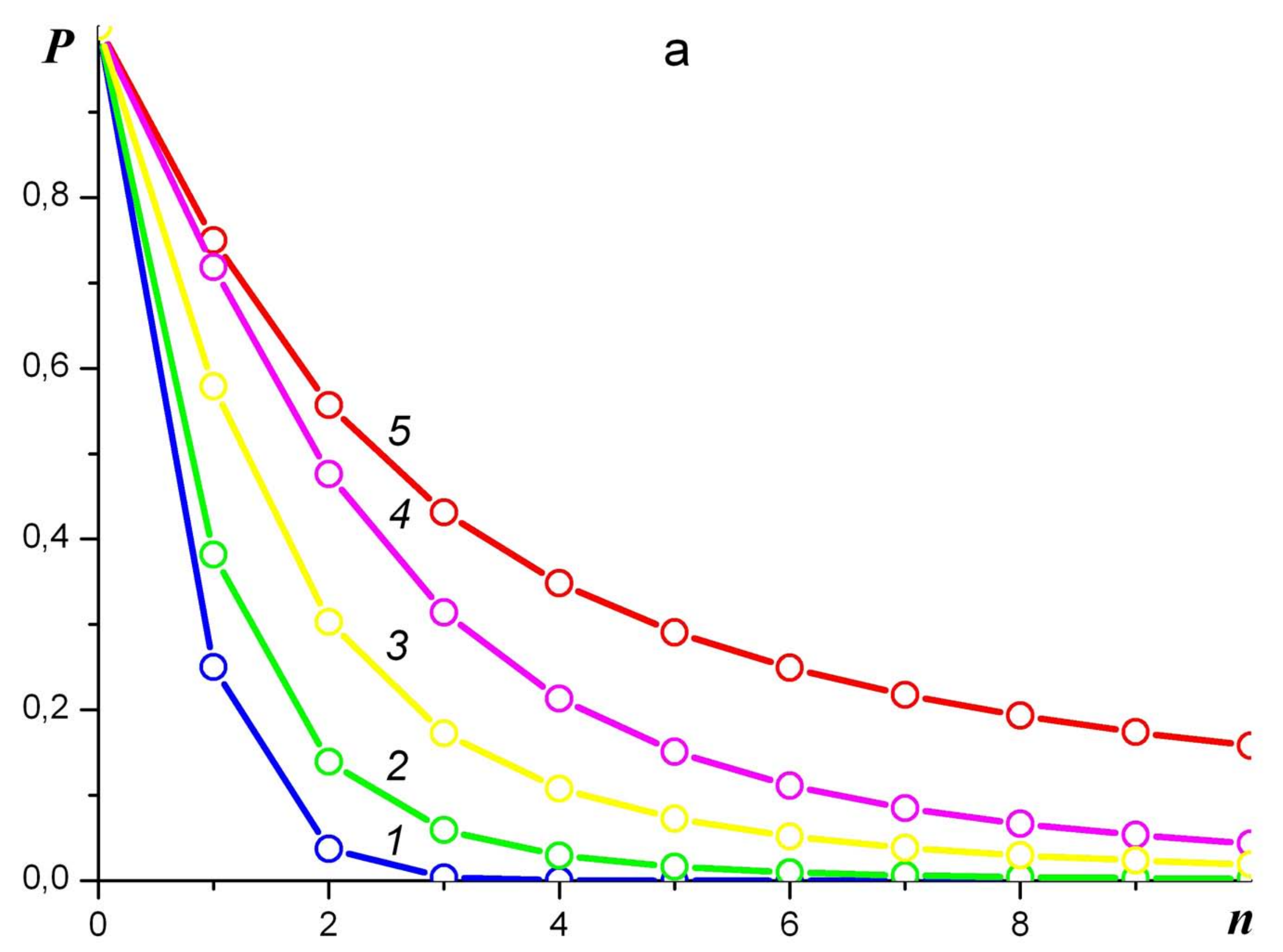}\\
 \includegraphics[width=80mm,angle=0]{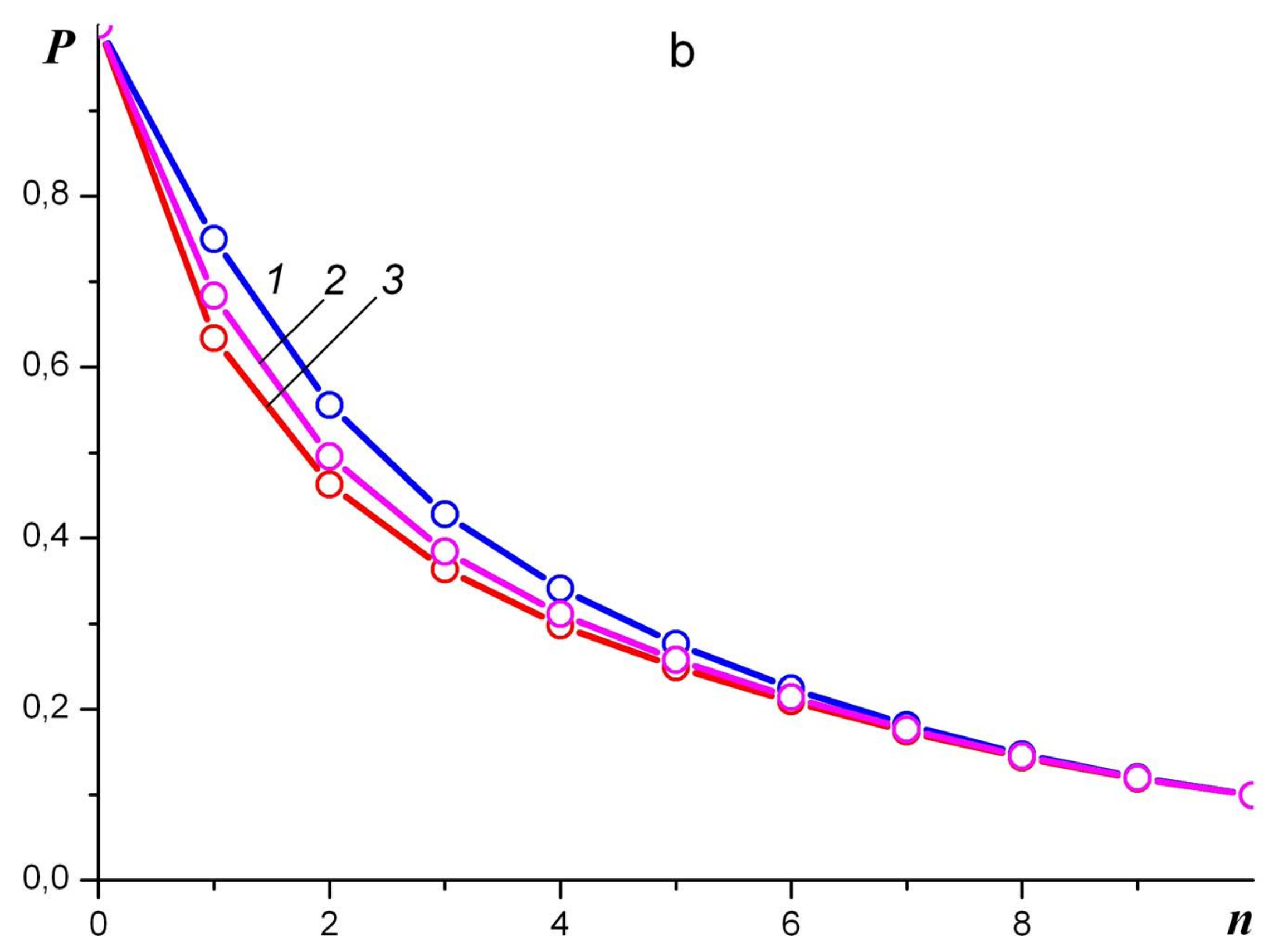}\\
 \caption{Creation probability of the regular hierarchical tree in dependence
of the whole number of its levels at: (a) $b=2$ and
$q=1.0001,1.5,1.9,1.99,1.9999$ (curves 1-5, respectively); (b)
$q=1.9999$, $n=10$ and $b=2,4,100$ (curves 1-3, respectively).}
\label{fig3}
\end{figure}

As numerical calculations show, the creation probability
(\ref{11}) takes meaningful values $P_0\leq1$ for deformation
parameters $q>1$ only. According to Fig. \ref{fig3}(a) the
dependence of this probability on the whole number of
tree levels has monotonically slowing down form whose decaying
rate decreases considerably only near the limit value $q=2$. On
the other hand, Fig. \ref{fig3}(b) shows that variation of the
branching index $b\gg1$ affects appreciably the dependence of the
the creation probability only for moderate numbers of tree levels
within the domain $2-q\ll1$.

\begin{figure}[!htb]
\centering
 \includegraphics[width=80mm,angle=0]{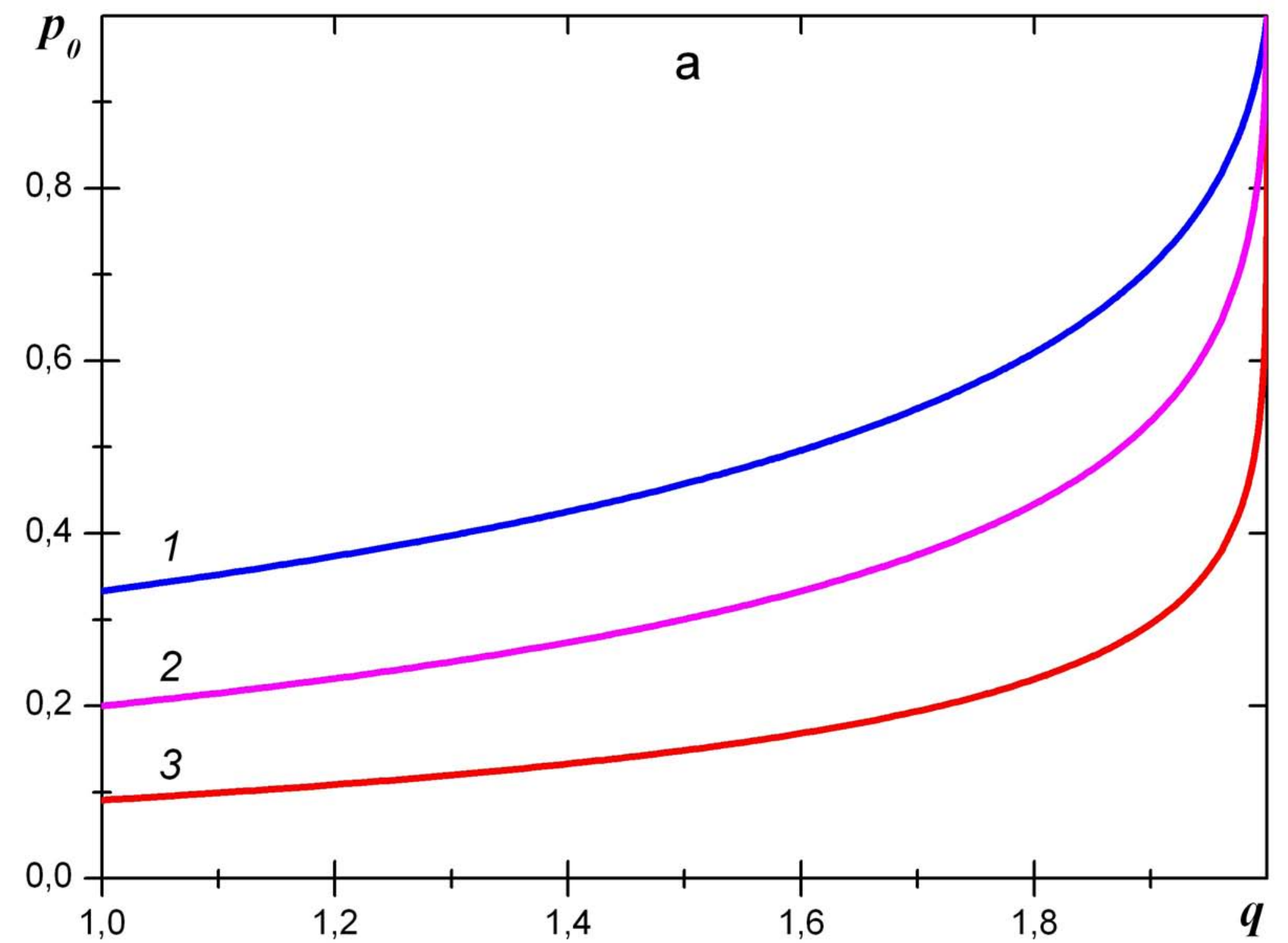}\\
 \includegraphics[width=80mm,angle=0]{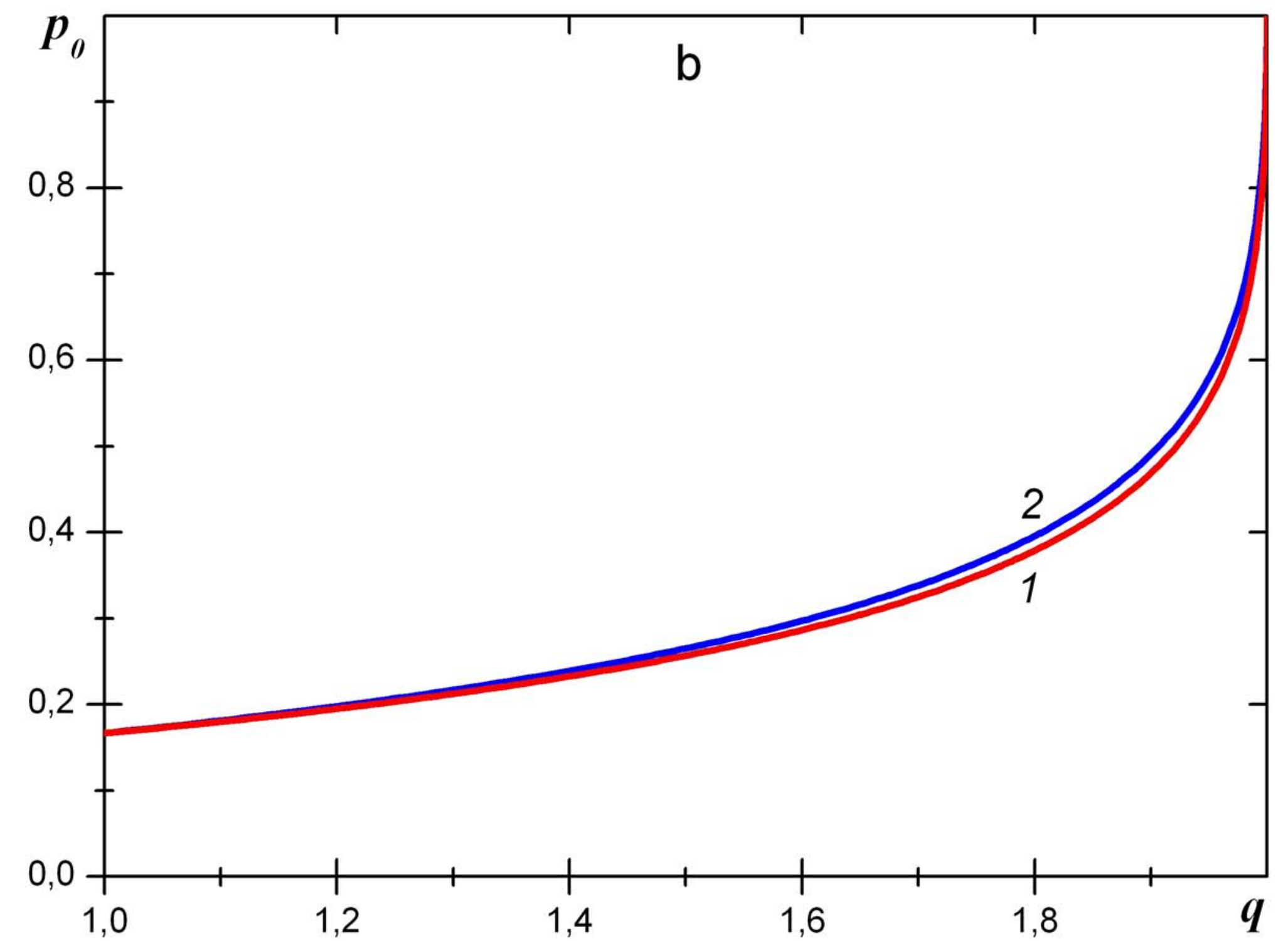}\\
  \caption{Top level probability of the regular tree as function of the deformation
  parameter at: (a) $b=2$ and $n=2,4,10$ (curves 1-3, respectively);
  (b) $n=5$ and $b=2,10^3$ (curves 1,2, respectively).} \label{fig4}
\end{figure}

Above data indicate distinctive feature in behavior of the regular
hierarchical tree near the limit value $q=2$ where the dependence
(\ref{aa}) has not any singularity. This feature is corroborated
with the dependence of the top level probability on the
deformation parameter depicted in Fig. \ref{fig4}. It is seen,
regardless of both total number of levels $n$ and
branching index $b$, this probability increases monotonically with
the $q$-growth to reach sharply the limit value $p_0=1$ in the
point $q=2$. Obviously, this means anomalous increasing
probabilities $p_l$ for the whole set of hierarchical levels (type
of shown in Fig. \ref{fig2}(a) with the curve 7). Though, within
the domain $2-q\ll1$, the ordinary normalization condition
$\sum_{l=0}^n p_l=1$ is violated appreciably, the definition
(\ref{77}) shows the deformed normalization condition (\ref{1a})
can be recovered at large parameter $q$. However, with overcoming
the border $q=2$ this condition is not satisfied at all. As a
result, we arrive at the conclusion that physically meaning values
of the deformation parameter are concentrated within the domain
$q\in[1,2]$.

\section{Degenerate tree}\label{Sec.4}

As shown in Fig. \ref{fig1}, the difference between regular and
degenerate trees is that {\it all} nodes multifurcate on each
level in the former case, while the {\it only one} node branches
in the latter. In this sense, the degenerate tree can be
considered as an antipode of the regular one to be studied
analytically.

According to Fig. \ref{fig1}(b), on the $l=1$ level, branching
process with index $b>1$ creates $N_1=b$ nodes with equal
probabilities $b^{-1}$. Next, on the $l=2$ level, $b-1$ nodes out
of $N_2=2(b-1)+1$ ones have the same probabilities $b^{-1}$, while
the $b$ rest nodes relate to the smaller value $b^{-2}$. On the
$l=3$ level, out of $N_3=3(b-1)+1$ nodes one has $b$ nodes with
probabilities $b^{-3}$, $b-1$ with $b^{-2}$ and $b-1$ with
$b^{-1}$. Hence, on the $l$ level $N_l=l(b-1)+1$ nodes are
partitioned into $l$ groups, among which $l-1$ ones contain $b-1$
nodes with probabilities $b^{-1}$, $b^{-2}$, $\dots$,
$b^{-(l-1)}$, while the last group has $b$ nodes with equal
probabilities $b^{-l}$. With accounting such a partitioning, the
creation probability of the $l$th hierarchical level is expressed
with the following relations:
\begin{eqnarray} \label{d}
%\begin{split}
\frac{p_l}{p_0}=\underbrace{\left[(b-1)\odot_q b^{-1}
\right]\oplus_q\dots\oplus_q\left[(b-1)\odot_q b^{-(l-1)}
\right]}_{l-1}\nonumber\\\oplus_q\left(b\odot_q b^{-l}\right)\nonumber\\
=\underbrace{\left[(b-1)\odot_q
b^{-1}\right]\oplus_q\dots\oplus_q\left[(b-1)\odot_q
b^{-l}\right]}_l\oplus_q
b^{-l}\nonumber\\:=\left[(b-1)\odot_q\left(S_{l+1}\ominus_q 1
\right)\right]\oplus_q b^{-l}.
%\end{split}
\end{eqnarray}
Here, in the last equation the sum of the deformed geometrical
series
\begin{equation}
 S_l:=\underbrace{1 \oplus_qb^{-1}\oplus_qb^{-2}\oplus_q\dots\oplus_q
b^{-(l-1)}}_{l}
 \label{e}
\end{equation}
is introduced. As shows related consideration in Appendix B, this sum is
expressed by the power series
\begin{equation}
 S_l=\sum\limits_{k=0}^{l-1}C^{k+1}_l(b)
 (1-q)^k b^{-\frac{k(k+1)}{2}}
 \label{S}
\end{equation}
with the deformed binomial coefficients \cite{QG}
\begin{equation}
  C^k_l(b)\equiv\prod_{m=0}^{k-1}\frac{1-b^{-(l-m)}}{1-b^{-(m+1)}}.
 \label{C}
\end{equation}
Inserting Eq. (\ref{S}) into the last relation (\ref{d}), one obtains the final
expression for the $l$th level creation probability
\begin{equation}
p_l=\frac{\left[1+(1-q)b^{-l}\right]\left[1+(1-q)\Sigma_l\right]^{b-1}-1}{1-q}p_0
 \label{Sigma6}
\end{equation}
where one denotes
\begin{equation}
 \Sigma_l\equiv S_{l+1}\ominus_q 1 =\frac{1 }{2-q}\sum\limits_{k=1}^{l}C^{k+1}_{l+1}(b)
 (1-q)^k b^{-\frac{k(k+1)}{2}}.
 \label{Sigma4}
\end{equation}

Within production representation
\begin{equation}
S_l=\frac{1}{1-q}\left\{\prod_{m=0}^{l-1}\Bigl[1+(1-q)b^{-m}\Bigl]-1\right\},
 \label{Ss2}
\end{equation}
one has
\begin{equation}
 \Sigma_l=
 \frac{1}{1-q}\left\{\frac{\prod_{m=0}^{l-1}\Bigl[1+(1-q)b^{-m}\Bigl]}{2-q}-1\right\}.
 \label{S4}
\end{equation}
Then, the probability (\ref{Sigma6}) takes the explicit form
\begin{equation}
p_l=\frac{\frac{1+(1-q)b^{-l}}{(2-q)^{b-1}}
\prod_{m=0}^{l-1}\Bigl[1+(1-q)b^{-m}\Bigl]^{b-1}-1}{1-q}p_0.
 \label{S6}
\end{equation}

In spite of apparent differences between the formulas (\ref{aa}) and
(\ref{S6}), direct calculations show actually coincident forms of the
probability distributions over hierarchical levels for both regular and
degenerate trees. Therefore, we postpone numerical study of the creation
probability for the degenerate tree before the following section where
consideration of the free-scale tree allows to compare all the results obtained
analytically.

\section{Free-scale tree}\label{Sec.5}

Above, we have considered two conceptual examples of hierarchical
trees with self-similar structure -- regular and degenerate trees
depicted in Fig. \ref{fig1}. In this section, we shall study a
free-scale tree whose structure is rather random, but the
probability distribution over hierarchical levels tends to the
power-law form inherent in self-similar statistical systems
\cite{OK}.

In this case, the probability distribution over tree levels is determined by
the discrete difference equation \cite{JETP_Let2000}
\begin{equation}
p_{l+1}-p_l=-p_l^q/\Delta,\qquad l=0,1,\dots,n
 \label{12a}
\end{equation}
accompanied with the deformed normalization condition (\ref{1a}) ($\Delta$
being a distribution dispersion). It is easily to show that in continual limit
$l\to\infty$ the equation (\ref{12a}) arrives at the power-law dependence
\cite{13a}
\begin{equation}
p_l=\left(p_0^{1-q}+\frac{q-1}{\Delta}~l\right)^{-\frac{1}{q-1}}
 \label{12}
\end{equation}
where the top level probability is
$p_0=\left(\frac{2-q}{\Delta}\right)^{\frac{1}{2-q}}$ for trees with total
number of levels $n\gg1$.

\begin{figure}[!htb]
\centering
\includegraphics[width=80mm,angle=0]{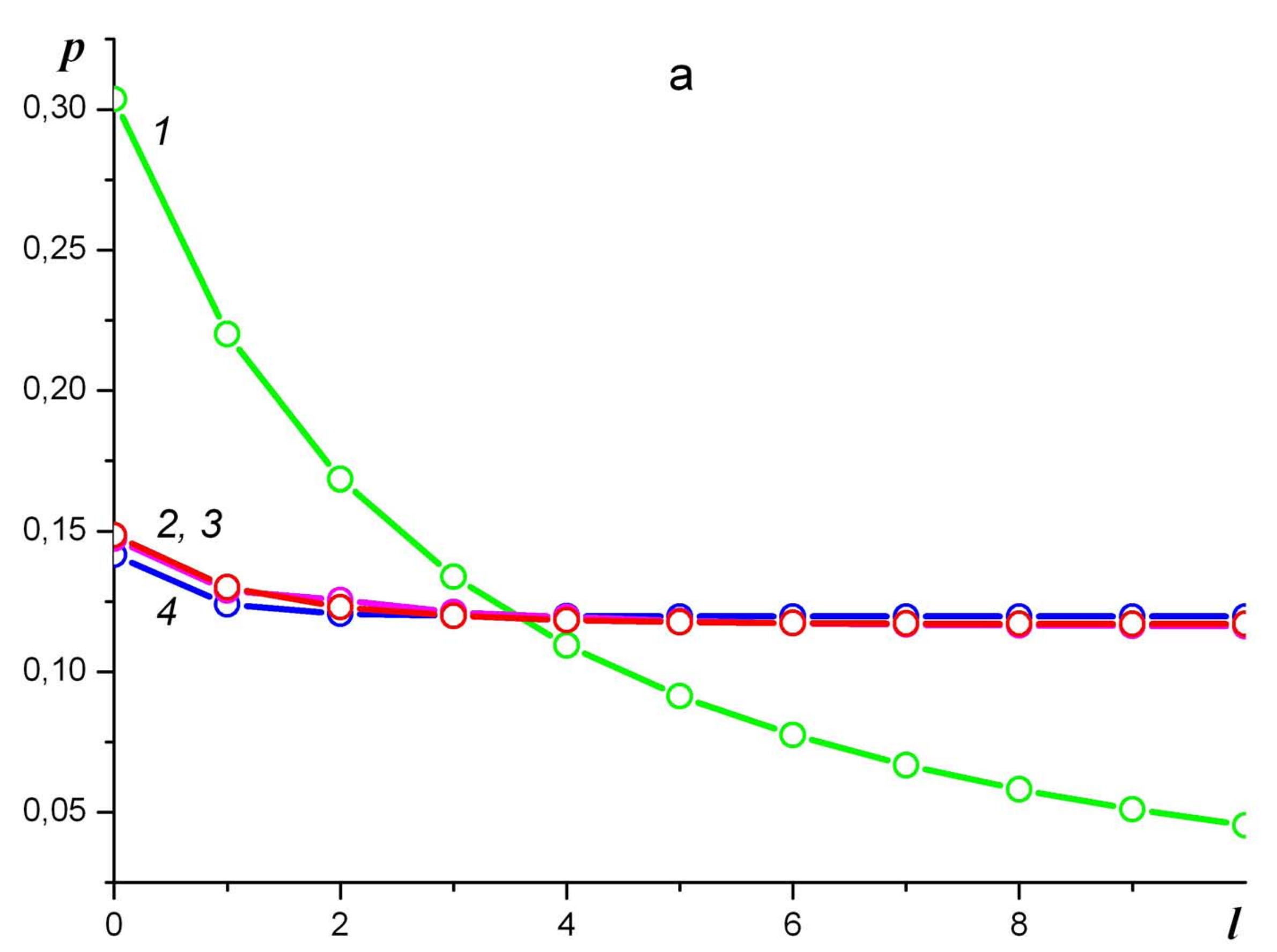}\\
\includegraphics[width=80mm,angle=0]{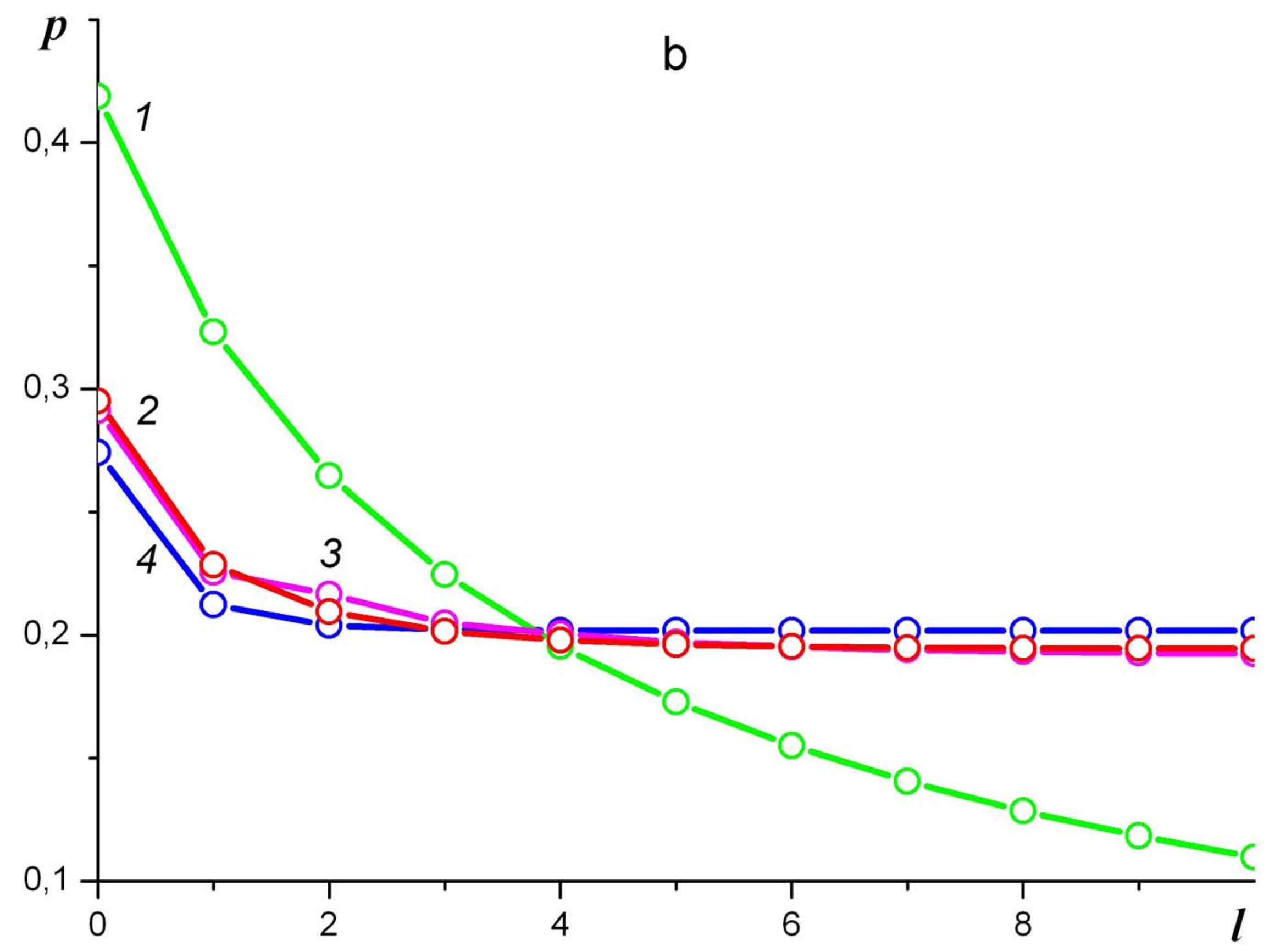}\\
\includegraphics[width=80mm,angle=0]{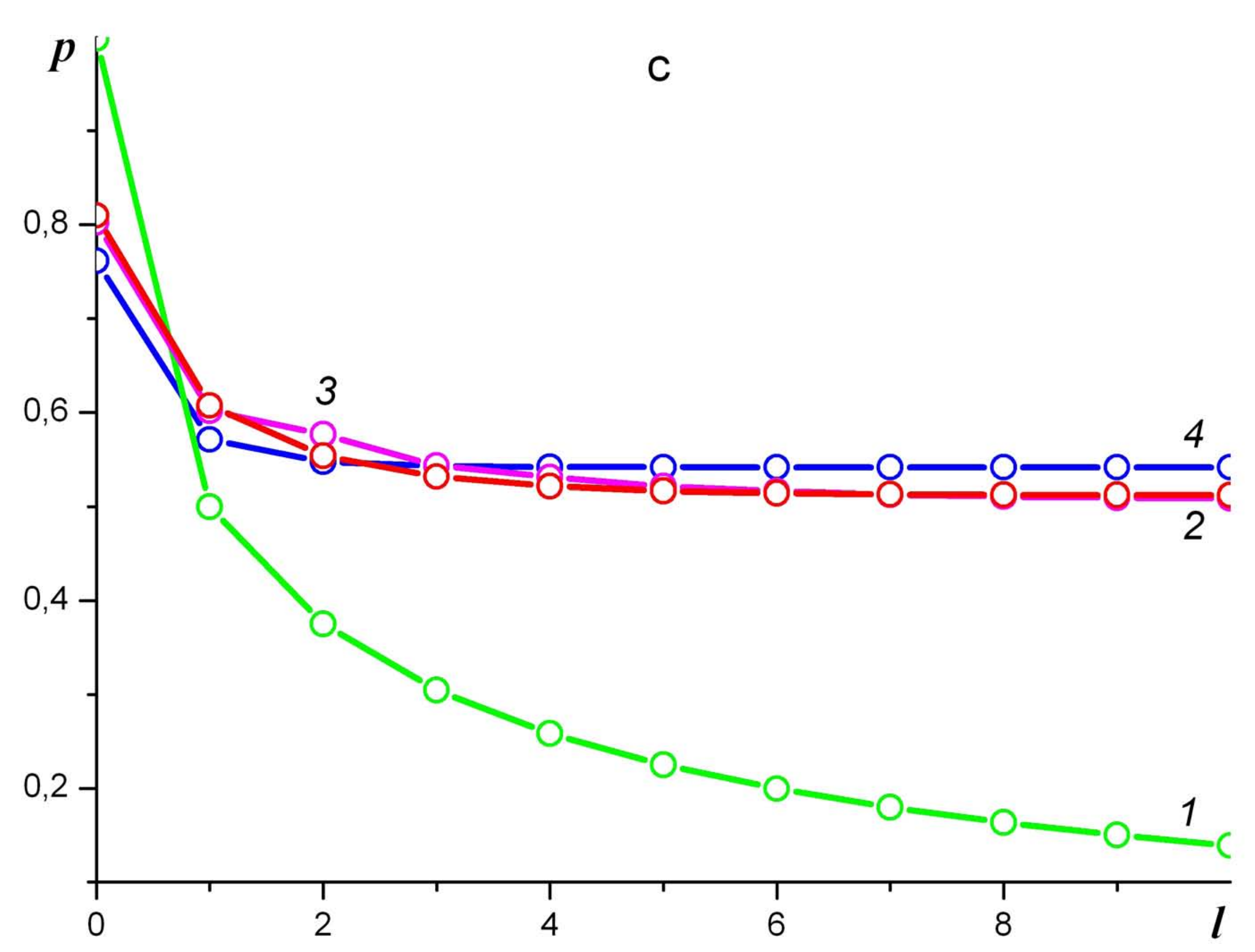}\\
 \caption{Probability distributions over hierarchical levels for free-scale,
 regular, Fibonacci and degenerate trees (curves 1-4, respectively) at
 $\Delta=2$, $b=2$, $n=10$ and $q=1.5$ (a), $q=1.9$ (b) and $q=1.9999$ (c).}
\label{fig5}
\end{figure}

\begin{figure}[!htb]
\centering
\includegraphics[width=80mm,angle=0]{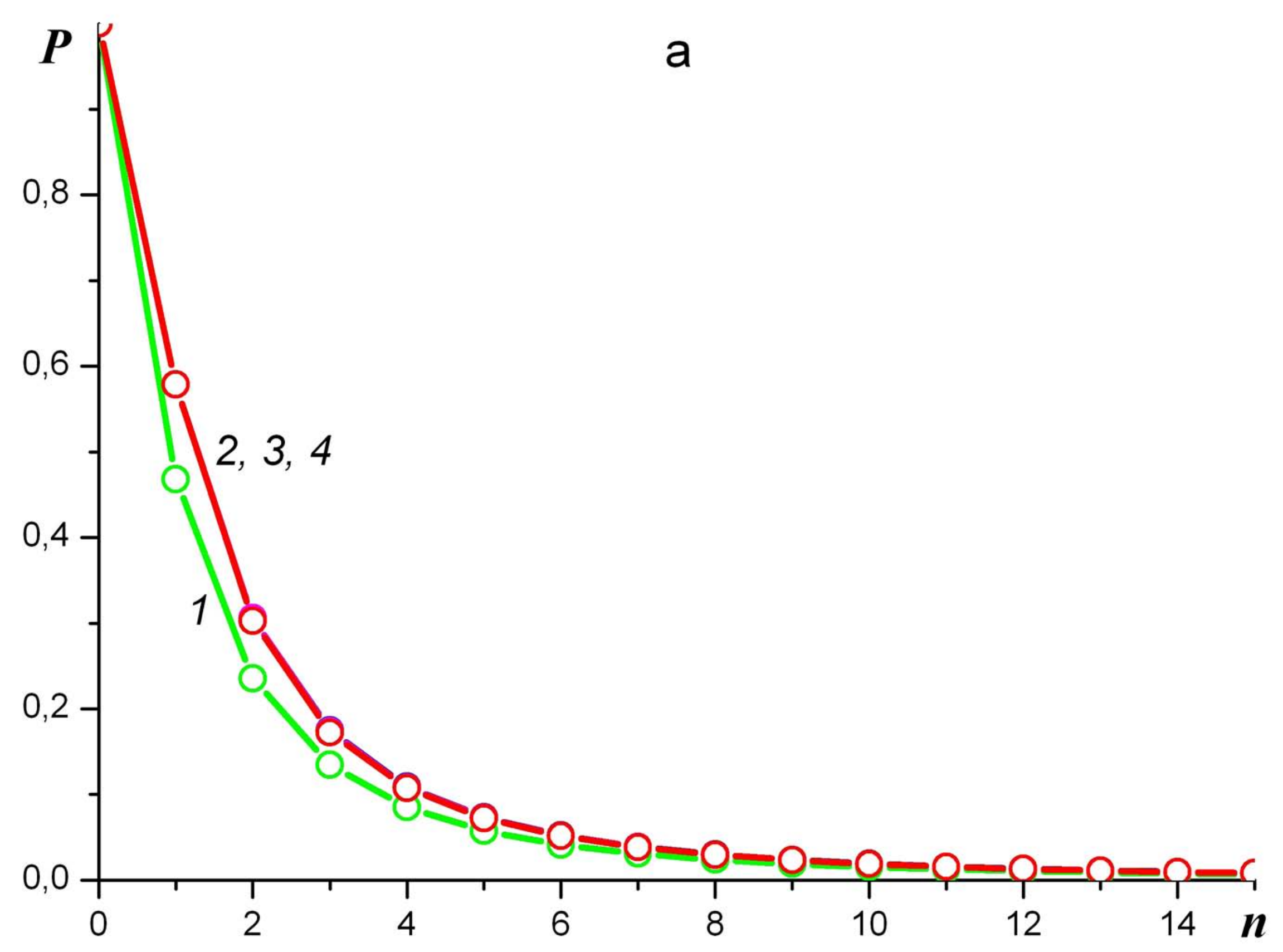}\\
\includegraphics[width=80mm,angle=0]{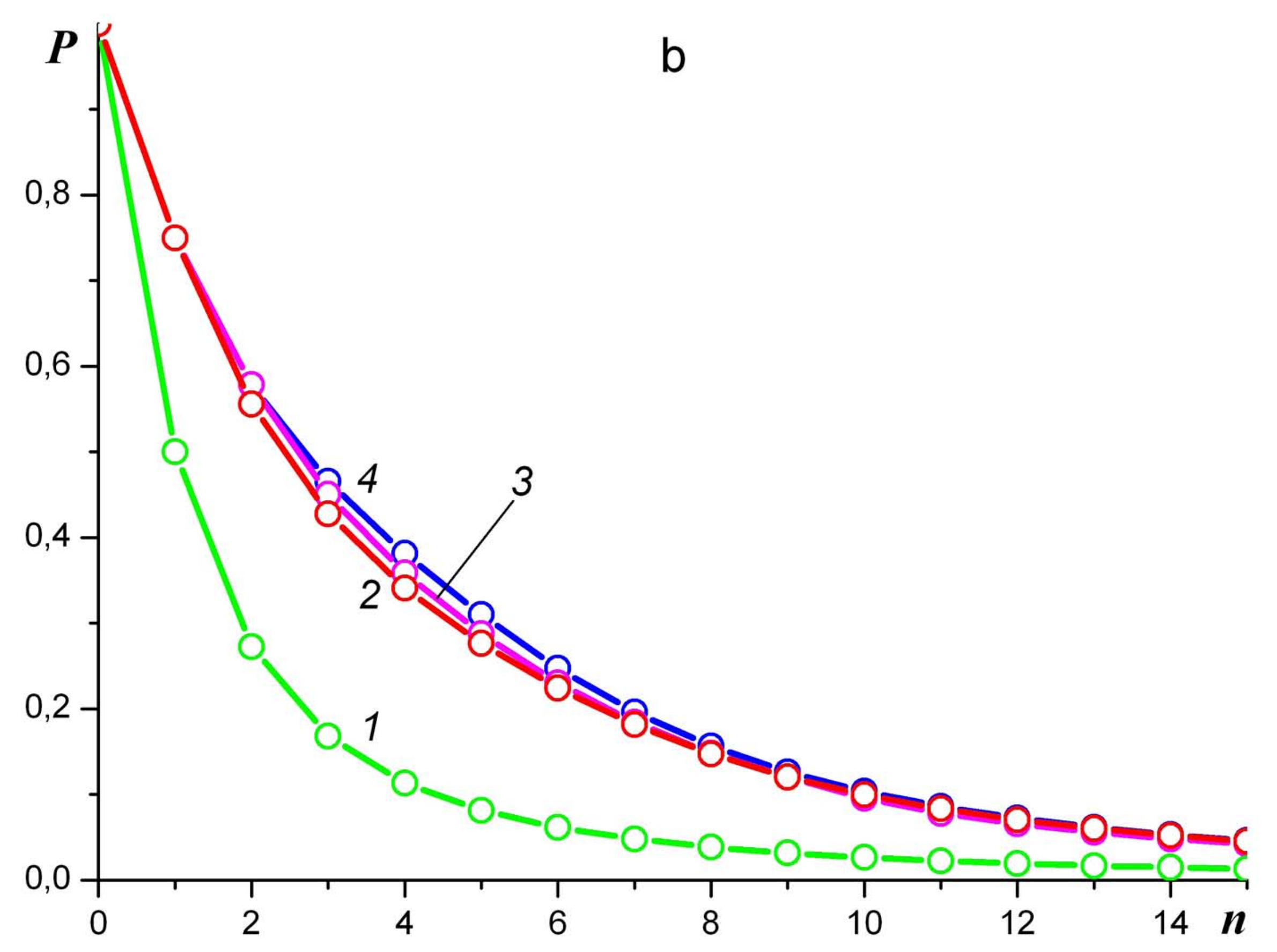}\\
\caption{Creation probabilities of free-scale, regular, Fibonacci
and degenerate hierarchical trees (curves 1-4, respectively) as
function of the whole level number at $\Delta=2$, $b=2$ and
$q=1.9$ (a) and $q=1.9999$ (b).} \label{fig6}
\end{figure}

In figures \ref{fig5} we compare the probability distributions
over
hierarchical levels of free-scale, regular and degenerate trees at
different values of the deformation parameter. It is seen at all
$q$-values the form of these distributions is actually equal for
regular and degenerate trees, but differs appreciably for
free-scale tree, where the level probability falls down much more
strong, than for both other trees. In accordance with such a
behavior, the creation probabilities depicted in Figs. \ref{fig6}
decays faster for the free-scale tree, than in the case of the
regular and degenerate ones. Characteristically, this difference
appears only within the domain $2-q\ll1$ of the deformation
parameter variation.

\begin{figure}[!h]
\centering
 \includegraphics[width=80mm,angle=0]{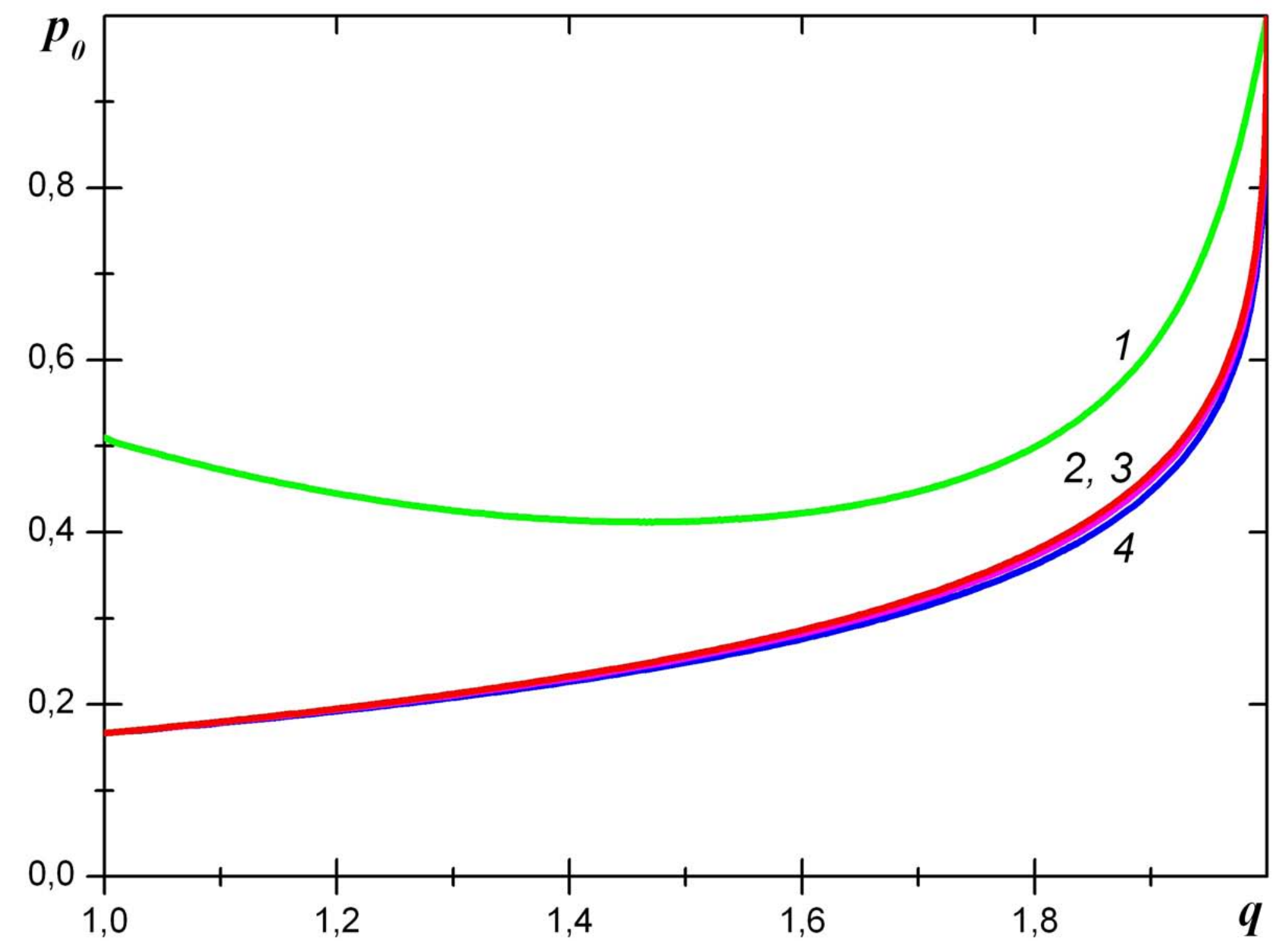}\\
  \caption{Top level probabilities for the free-scale, regular,
  Fibonacci and degenerate trees
  (curves 1-4, respectively) as function of the deformation parameter at
  $\Delta=2$, $b=2$ and $n=5$.} \label{fig7}
\end{figure}

As shown in the end of the section \ref{Sec.3}, such a behavior is
stipulated by the singular dependence of the top level probability
$p_0$ on the deformation parameter near the point $q=2$. According
to Fig. \ref{fig7} this
singularity is inherent in all considered hierarchical trees.

\section{Arbitrary tree}\label{Sec.6}

\begin{figure}[!htb]
\centering
\includegraphics[width=90mm,angle=0]{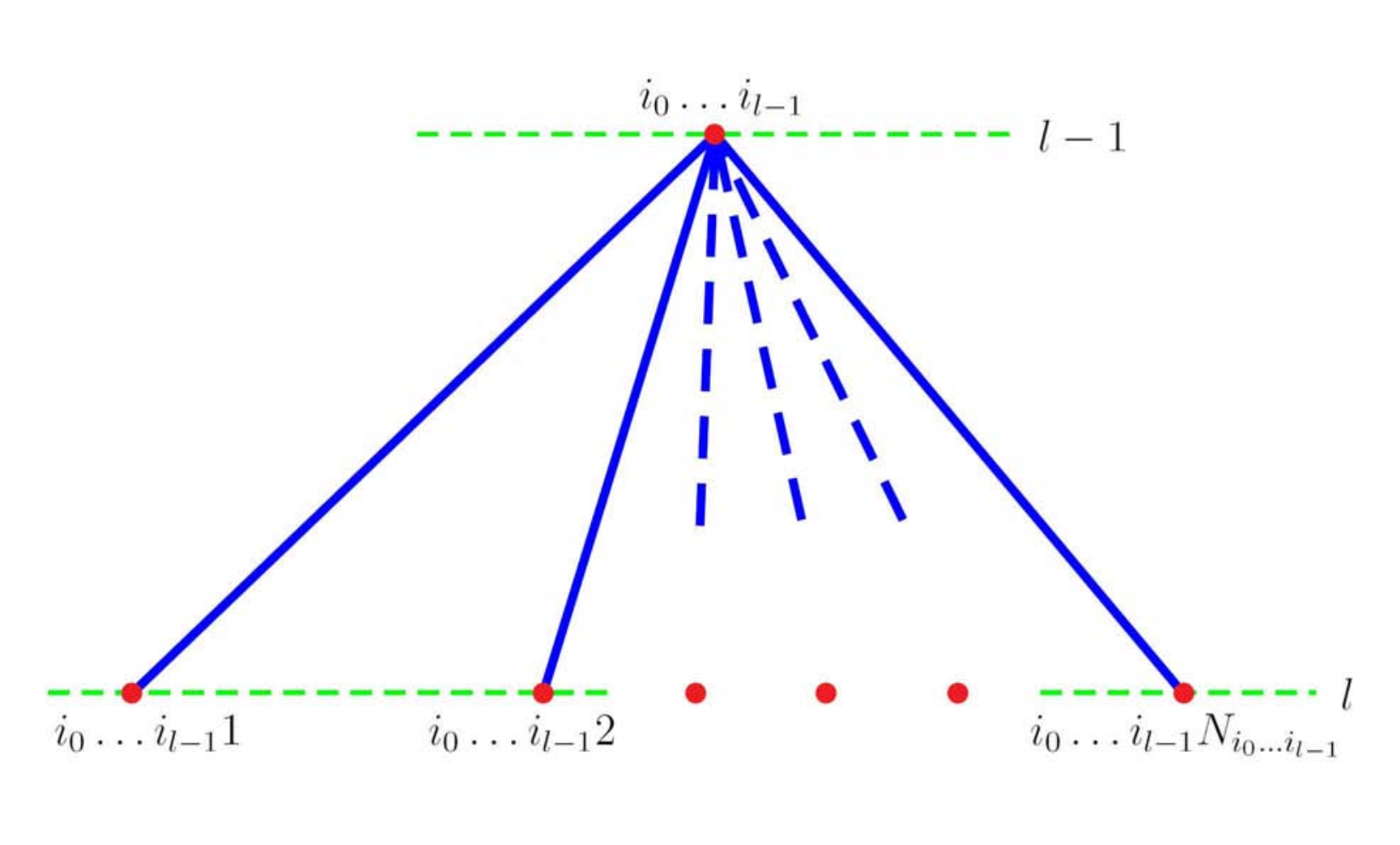}
\caption{Node parametrization within a hierarchical cluster.}
\label{fig8}
\end{figure}

Now, we are in position to consider an arbitrary hierarchical
tree, over whose levels $l=0,1,\dots,n$, $n\geq1$ are distributed
$N_l$ nodes $i_0 i_1\dots i_l$ with the probabilities
$p_{i_0i_1\dots i_l}$ \footnote{In accordance with
Ref.\cite{Rammal}, a node coordinate of a hierarchical tree
represents so names {\it $p$-adic number} $i_0i_1\dots i_n$ where
the first digit $i_0=1$ relates to the major ancestor on the
uppermost level $l=0$, the second $i_1$ numbers its sons on the
lower level $l=1$, and so on -- up to the last digit $i_n$
numbering the lowest descendants on the bottom level $l=n$.}. The
main peculiarity of hierarchical trees is known to be a clustered
structure, whose fragment is depicted in Fig. \ref{fig8}: nodes
$i_0\dots i_{l-1}i_l$ of the $l$-level form a
cluster $i_0\dots i_{l-1}$ on the $(l-1)$-level; in turn, clusters $i_0\dots
i_{l-2}i_{l-1}$ form supercluster $i_0\dots i_{l-2}$ on the following level
$l-2$, et cetera. Above clustering process spreads over upper levels $l-3$,
$l-4,\dots$ up to the pair of the top levels $l=1$ and $l=0$ where $N_{i_0}$
nodes $i_1$ form the superior node $i_0$. Along this way, the node
probabilities on hierarchical levels ranged bottom-up are as follows:
$p_{i_0\dots i_{n-1}i_n}$, $p_{i_0\dots i_{n-1}}$, $\dots,$ $p_{i_0 \dots
i_l}$, $\dots,$ $p_{i_0i_1}$, $p_{i_0}\equiv p_0$. Let us calculate these
probabilities considering hierarchical levels top-down.

On the uppermost level $l=0$, one has a single node $i_0=1$ related to the
probability $p_0\equiv p_{i_0}$. With passage down to the level $l=1$, this
node multifurcates into a cluster comprising of $N_{i_0}$ nodes $i_1$. Because
of the identity of this nodes, they are characterized by the equal
probabilities
\begin{equation} \label{15a}
p_{i_0i_1}=p_0 N_{i_0}^{-1}.
\end{equation}
In similar manner, on the following level $l=2$ one obtains the node
probabilities
\begin{equation} \label{15b}
p_{i_0i_1i_2}=p_{i_0i_1}N_{i_0i_1}^{-1}=p_0\left(N_{i_0}N_{i_0i_1}\right)^{-1}.
\end{equation}
Iteration of this procedure down to an arbitrary level $l$ yields the required
result
\begin{equation} \label{15c}
p_{i_0\dots i_l}=p_0\left(\prod\limits_{m=0}^{l-1}N_{i_0\dots i_m}\right)^{-1}
\end{equation}
where $N_{i_0\dots i_m}$ is the node number within the cluster $i_0\dots i_m$.

Generalization of the first equality (\ref{a}) arrives at the expression of the
creation probability of an arbitrary level $l$ through a set of related node
probabilities. This expression is reduced to the following $l$-fold deformed
sum:
\begin{equation} \label{15v}
\frac{p_l}{p_0}=\biguplus\limits_{i_1=1}^{N_{i_0}}\dots
\biguplus\limits_{i_l=1}^{N_{i_0\dots i_{l-1}}}\frac{p_{i_0\dots
i_l}}{p_0},\quad l\ne0.
\end{equation}
Respectively, the normalization condition (\ref{1a}) takes the form
\begin{equation} \label{15w}
p_0\oplus_q p_0\biguplus\limits_{i_1=1}^{N_{i_0}}
\dots\biguplus\limits_{i_{n}=1}^{N_{i_0\dots i_{n-1}}} \frac{p_{i_0\dots
i_{n}}}{p_0}=1.
\end{equation}
Above, we have used the notation of the deformed sum of $n$ terms:
\begin{equation} \label{15u}
\biguplus\limits_{i=1}^{n}a_i\equiv a_1\oplus_q a_2\oplus_q\dots\oplus_qa_n.
\end{equation}

It is worth noting the characteristic peculiarity of above consideration: the
node probabilities (\ref{15c}) are determined with making use of non-deformed
algebra, while the definition (\ref{15v}) of the level probabilities $p_l$ is
based on the use of deformed summation (\ref{15u}). A ground of such a
partitioning is that the former of these probabilities relates to the
configuration of hierarchical trees, while the latter describes their
statistical properties.

\begin{figure}[!h]
\centering
 \includegraphics[width=80mm,angle=0]{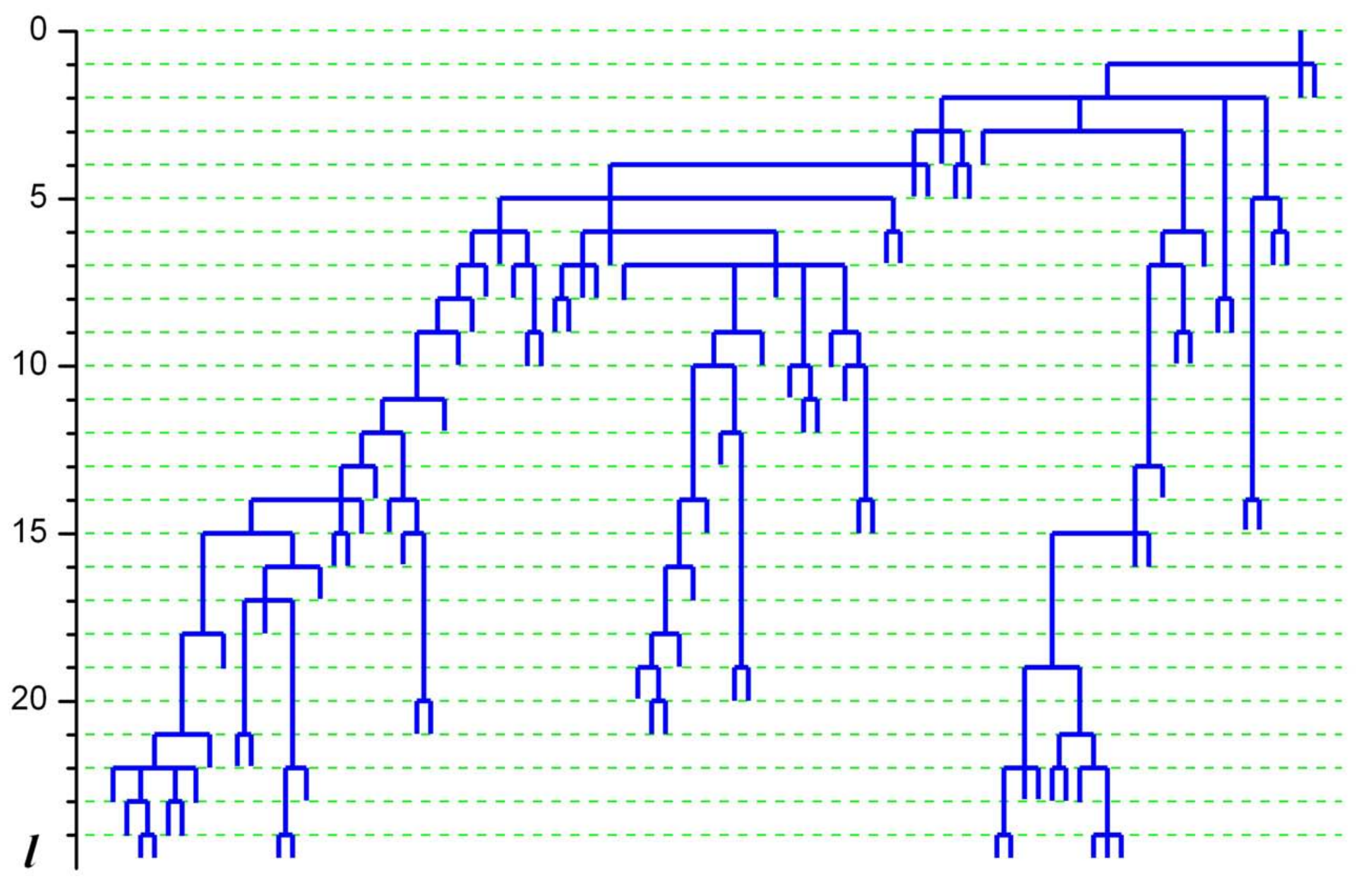}\\
  \caption{Schematic representation
  of evolution tree (from Ref. \cite{Science}).} \label{fig9.pdf}
\end{figure}

\begin{figure}[!h]
\centering
 \includegraphics[width=85mm,angle=0]{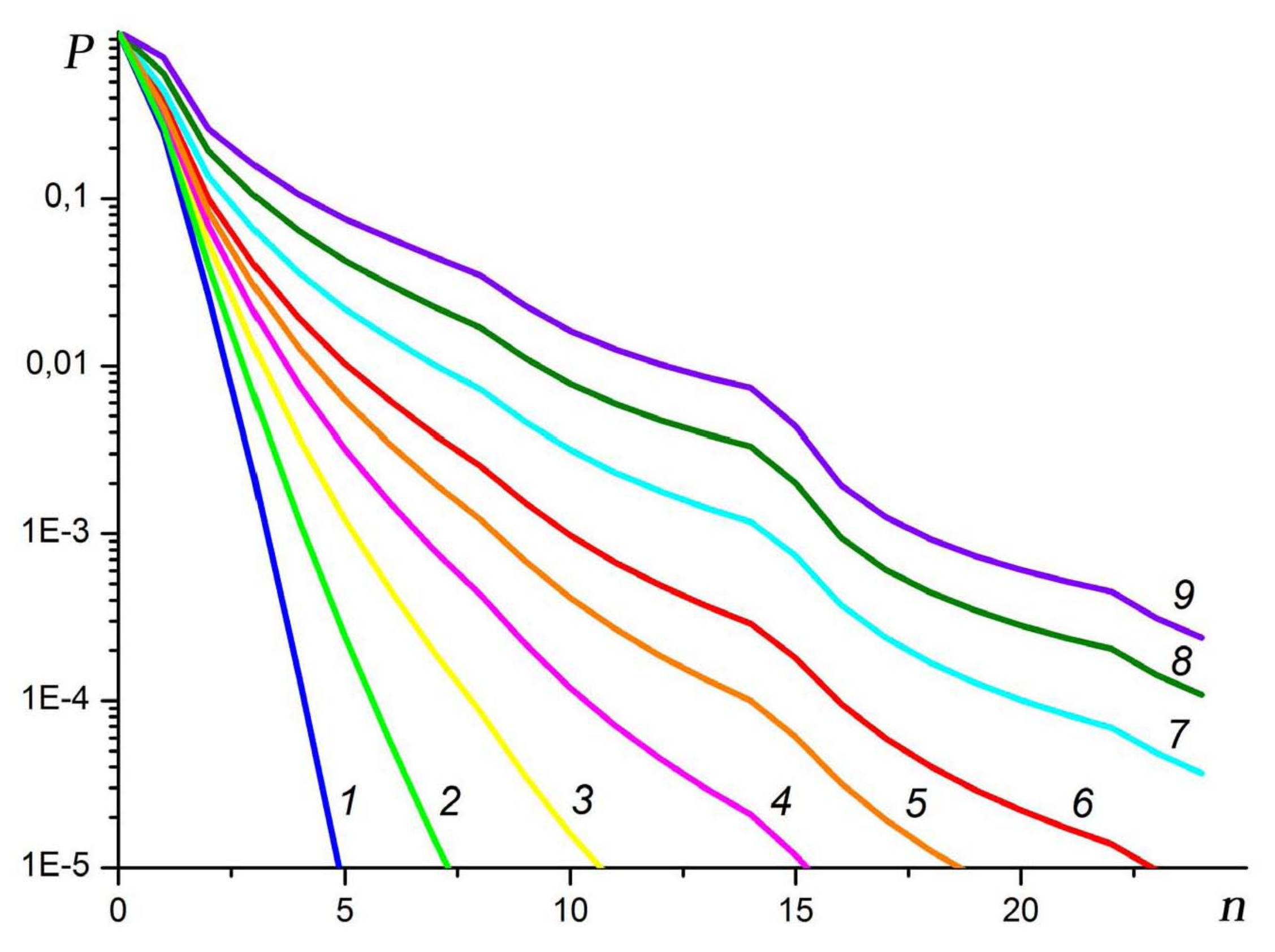}\\
  \caption{Creation probability of the evolution tree vs. the level number at:
  $q = 1.0001, 1.1, 1.2, 1.3, 1.4, 1.5, 1.7, 1.9, 1.9999$ (curves
1-9, respectively).} \label{fig10.pdf}
\end{figure}

In conclusion, we consider two examples of applying above theory,
among which the former concerns the Fibonacci tree (Fig.
\ref{fig1}(b)), while the latter relates to the schematic
evolution tree shown in Fig. \ref {fig9.pdf}
(in the last case, nodes identify substantial stages in evolution
of life, e.g., human is situated on the 24th level). Using the
formulas (\ref {15c}) and (\ref {15v}) for the node and level
probabilities, obeying the normalization condition (\ref{15w}), we
show that probability distributions of the Fibonacci tree depicted
in Figs. \ref{fig5} - \ref{fig7} does not differ actually from
related dependencies for both regular and degenerate trees. What
about the evolution tree, its probability distributions (Fig.
\ref{fig10.pdf})
show that a presence of the stopped branches (type of two
rightmost ones in Fig. \ref {fig9.pdf}) considerably decreases
creation probability of new hierarchical level. Particularly, the
probability of human appearance takes values more than $10^{-4}$
only at the deformation parameter $q=1.9999$.

\section{Concluding remarks}\label{Sec.7}

To escape ambiguities we are worthwhile to stress that our consideration
concerns rather the probabilistic picture of creation of hierarchical trees
themselves, than hierarchical phenomena and processes evolving on these trees
(for example, hierarchically constrained statistical ensembles \cite{13a},
diffusion processes on multifurcating trees \cite{Huberman}, et cetera). Among
others we have studied analytically both regular and degenerate trees to
confirm the coincidence of both analytical and numerical results following from
the developed scheme being applicable to an arbitrary tree.

A principle peculiarity of the probabilistic picture elaborated is a
partitioning deformed and non-deformed values. So, effective energies of
hierarchical levels in Eq. (\ref{2}) are non-deformed values because creation
of hierarchical structure does not break the conservation law of the energy
being additive value. Moreover, the node probabilities are determined with
making use of non-deformed relation (\ref{15c}) because these probabilities
relate to the configuration of hierarchical tree itself (in other words, they
are determined by geometrical, but not probabilistic reasons). At the same
time, the hierarchy appearance deforms essentially the probability relations
(\ref{1a}), (\ref{15w}), (\ref{5}), (\ref{11}) and (\ref{13}) due to coupling
level probabilities $p_l$. Similarly, the definition (\ref{15v}) of these
probabilities through corresponding node values is based on the use of deformed
summation (\ref{15u}).

Making use of deformed algebra shows increase of probabilities $p_l$ for the
whole set of hierarchical levels to take anomalous character near the point
$q=2$. The deformed normalization condition (\ref{1a}) is fulfilled only at
$q\leq 2$, while it is broken with overcoming the border $q=2$. As a result,
physically meaning values of the deformation parameter belong to the domain
$q\in[1,2]$.

Comparison of the probability distributions over hierarchical
levels of free-scale, regular, Fibonacci and degenerate trees
shows (Fig. \ref{fig5}) the form of these distributions differs
appreciably at all $q$-values only for free-scale tree where the
level probability falls down much more strong. In accordance with
such a behavior, the creation probabilities depicted in Fig.
\ref{fig6} decays faster for the free-scale tree, than for the
rest ones. Characteristically, this difference appears within the
condition $2-q\ll1$ only.

Expression (\ref {15c}) -- (\ref {15w}) and (\ref {11}) are a basis for
numerical studies of arbitrary hierarchical structures, for example complex
defect structures of solids subject to intensive external influence type of
rigid radiation treatment. Unlike the amorphous systems, the number of
structure levels of a real crystal is rather not large: usually, among
different spatial scales, it is accepted to distinguish micro-, meso- and
macroscopic levels \cite{NOVA}. To study a real structure, one needs first to
distribute the whole ensemble of defects over hierarchical levels $l=0,1,
\dots, n$; then, one calculates on each of them a number of defects $N _
{i_0i_1\dots i _ {l-1}} $ belonging to the cluster $i_0i_1\dots i _ {l-1} 1$,
$i_0i_1\dots i _ {l-1} 2$, $ \dots $, $i_0i_1\dots i _ {l-1} N_{i_1i_2\dots i _
{l-1}} $ and attributes the probability $p_{i_0\dots i_l}$ to this cluster in
accordance with Eq. (\ref {15c}). Next, the level probabilities $p_l$ are
calculated according to definition (\ref {15v}) where the top value $p_0$ is
fixed by the normalization condition (\ref {15w}). Finally, the creation
probabilities $P_n$ of hierarchical trees are determined by the equality (\ref
{11}).

%\section{Acknowledgements}\label{sec:level7}

%We are grateful to anonymous referees for constructive criticism.

\appendix
\section{Main rules of deformed algebra}\label{Sec.}

Following \cite{Borges}, let us present the main equations of the deformed
algebra. Related formalism is known to be based on the generalized definition
of the logarithm and exponential functions
\begin{eqnarray} \label{6}
%\begin{split}
\ln_q(x):= \frac{x^{1-q}-1}{1-q},\nonumber\\ \exp_q(x):=
\left[1+(1-q)x\right]_+^{1\over 1-q}
%\end{split}
\end{eqnarray}
being characterized by a deformation parameter $q\geq 0$ with the notion
$[y]_+\equiv \max(0,y)$. For some numbers $x,y>0$, deformed product and ratio
are defined with the following relations:
\begin{eqnarray} \label{7}
%\begin{split}
x\otimes_q y:=\left[x^{1-q}+y^{1-q}-1\right]_+^{1\over
1-q},\nonumber\\ x\oslash_q
y:=\left[x^{1-q}-y^{1-q}+1\right]_+^{1\over 1-q}.
%\end{split}
\end{eqnarray}
Respectively, deformed sum and difference read
\begin{eqnarray} \label{77}
%\begin{split}
x\oplus_q y:=x+y+(1-q)xy,\nonumber\\ x\ominus_q
y:=\frac{x-y}{1+(1-q)y}
%\end{split}
\end{eqnarray}
where the condition $y\ne-\frac{1}{1-q}$ is implied. The $n$-fold deformed sum
of identical terms is defined as follows:
\begin{equation}
n\odot_q x\equiv\underbrace{x\oplus_qx\oplus_q\dots\oplus_q
x}_{n}:=\frac{[1+(1-q)x]_+^n-1}{1-q}.
 \label{77a}
\end{equation}
The rules (\ref{7}), (\ref{77}) ensure the following properties of the
$q$-logarithm and the $q$-exponential (\ref{6}):
\begin{eqnarray} \label{8}
%\begin{split}
\ln_q(x\otimes_q y)=\ln_q x+\ln_q y,\nonumber\\\ln_q(x\oslash_q y
)=\ln_q x-\ln_q y;\nonumber\\
\exp_q(x)\otimes_q\exp_q(y)=\exp_q(x+y),\nonumber\\
\exp_q(x)\oslash_q\exp_q(y)=\exp_q(x-y).
%\end{split}
\end{eqnarray}

\section{Deformed sum of terms of a geometrical progression}

Let a geometrical sequence $a, ar, ar^2,\dots, ar^{n-1}$ is determined by the
common ratio $r$, the scale factor $a$ and the term number $n$. Within deformed
summation rule (\ref{77}), direct calculations at lower numbers $n=2,3,\dots$
show the sum of terms of a geometrical progression
\begin{equation}
S_n:=\underbrace{a\oplus_q ar\oplus_q\oplus_q ar^2\oplus_q\dots\oplus_q
ar^{n-1}}_n
 \label{a1}
\end{equation}
can be written as the series
\begin{equation}
S_n:=a\sum\limits_{m=0}^{n-1}\sigma_n^m[(1-q)a]^m
 \label{Sigma}
\end{equation}
with unknown coefficients $\sigma_n^m$. Iteration of Eq. (\ref{a1}) yields the
chain of the following relations:
\begin{eqnarray}
%\begin{split}
 S_{n+1}:=S_{n}\oplus_q(ar^n)=\left(S_{n}+ar^n\right)+(1-q)
 S_{n}(ar^n)\nonumber\\=\left\{a\left(\sigma_{n}^0+r^n\right)+
 a\sum\limits_{m=1}^{n-1}\sigma_n^m[(1-q)a]^m\right\}\nonumber\\
 +(1-q)a(ar^n)\sum\limits_{m=0}^{n-1}\sigma_n^m[(1-q)a]^m\nonumber\\
 =a\sigma_{n+1}^0+
 a\sum\limits_{l=0}^{n-2}\sigma_n^{l+1}[(1-q)a]^{l+1}\nonumber\\
 +a[(1-q)a]r^n\sum\limits_{m=0}^{n-1}\sigma_n^m[(1-q)a]^m\nonumber\\
 =a\sigma_{n+1}^0+
 a\sum\limits_{l=0}^{n-2}\left(\sigma_n^{l+1}+r^n\sigma_n^l\right)[(1-q)a]^{l+1}\nonumber\\
 +a[(1-q)a]r^n\sigma_n^{n-1}[(1-q)a]^{n-1}\nonumber\\
 =a\sigma_{n+1}^0+
 a\sum\limits_{m=1}^{n-1}\left(\sigma_n^{m}+r^n\sigma_n^{m-1}\right)[(1-q)a]^{m}\nonumber\\
 +a\sigma_n^{n-1}[(1-q)a]^{n}r^n.
%\end{split}
\end{eqnarray}
Here, the in the first line takes into account the definition
(\ref{77}); in the second line, the series (\ref{Sigma}) is
applied to single out the term related to $m=0$ within the braces;
in the fourth line, the first term is written in accordance with
the definition (\ref{Sigma}) related to the term $m=0$, while the
summation index $l=m-1$ is introduced in the second term; in the
sixth line, the second term contains both sums over $l$ and $m$ of
the previous line, the last term relates to the index $m=n-1$; in
the eighth line, we return to the summation index $m=l+1$. As a
result, the series (\ref{Sigma}) takes the form
\begin{eqnarray}
 S_{n}=a s_{n}+
 a\sum\limits_{m=1}^{n-2}\left(\sigma_{n-1}^{m}
 +r^{n-1}\sigma_{n-1}^{m-1}\right)[(1-q)a]^{m}\nonumber\\
 +a\sigma_{n-1}^{n-2}[(1-q)a]^{n-1}r^{n-1}
 \label{ab6}
\end{eqnarray}
where the sum
\begin{equation}
 s_n\equiv\sum\limits_{m=0}^{n-1}r^m=\frac{1-r^n}{1-r}
 \label{a3}
\end{equation}
of the ordinary geometrical progression $1, r, r^2,\dots, r^{n-1}$ was used.
Comparison of the terms of Eqs. (\ref{Sigma}) and (\ref{ab6}) related to the
equal $m$ indexes arrives at the following iteration relations:
\begin{eqnarray}
 \sigma_n^0=s_n;\label{b1}
\\ \sigma_n^m=\sigma_{n-1}^{m}+\sigma_{n-1}^{m-1}r^{n-1},\quad
 m\in[1,n-2];\label{b2}\\
 \sigma_n^{n-1}=\sigma_{n-1}^{n-2}r^{n-1}.
\label{b3}
\end{eqnarray}

The first of these terms gives the explicit expression of the lowest power
coefficient in the series (\ref{Sigma}). It is easily to convince the
regression (\ref{b2}) is satisfied with the insertion
\begin{equation}
 \sigma_n^{m}=\sum\limits_{l=0}^{n-1}\sigma_l^{m-1} r^l
 \label{ab3}
\end{equation}
whose iteration yields
\begin{eqnarray}
 \sigma_n^m=\sum\limits_{l=0}^{n-1}\sigma_l^{m-1} r^l
 =\sum\limits_{l=0}^{n-1}\sum\limits_{k=0}^{l-1}\sigma_{k}^{m-2}
 r^{l+k}=\dots\nonumber\\
 =\sum\limits_{l_{m-1}=0}^{n-1}r^{l_{m-1}}
 \sum\limits_{l_{m-2}=0}^{l_{m-1}-1}r^{l_{m-2}}
 \dots\sum\limits_{l_0=0}^{l_{1}-1}\sigma_{l_{0}}^0 r^{l_{0}}.
\end{eqnarray}
However, the last expression is inconvenient for direct calculations because it
contains connected exponents of the ratio $r$ with the upper limits of the
consequent sums. Hence, let us calculate explicitly the coefficients
(\ref{ab3}) for small indexes $m$:
\begin{eqnarray}
%\begin{split}
 \sigma_{n}^1=\sum_{l=0}^{n-1}\sigma_{l}^{0}r^l=\sum_{l=0}^{n-1}s_{l}r^l
 =\sum_{l=0}^{n-1}\frac{1-r^n}{1-r}r^l\nonumber\\=r\frac{(1-r^n)(1-r^{n-1})}{(1-r)(1-r^2)},\nonumber\\
 \sigma_{n}^2=\sum_{l=0}^{n-1}\sigma_{l}^{1}r^l=r\sum_{l=0}^{n-1}\frac{(1-r^{l})(1-r^{l-1})}
 {(1-r)(1-r^2)}r^l\nonumber\\=r^3\frac{(1-r^n)(1-r^{n-1})(1-r^{n-2})}{(1-r)(1-r^2)(1-r^3)},\nonumber\\\dots
%\end{split}
\end{eqnarray}
These expressions show that above coefficients are proportional to
the fractions, whose denominators represent the production of the
terms $1-r^{l+1}$ with growing powers $l=0,1,\dots$, while
numerators contain the same number of the terms $1-r^{n-l}$ with
dropping powers. As a result, we suppose the coefficients to be
found in the following form:
\begin{eqnarray} \label{ca3}
%\begin{split}
 \sigma_{n}^m=r^{\sum_{k=1}^{m}k}\prod_{l=0}^{m}\frac{1-r^{n-l}}
 {1-r^{l+1}}\nonumber\\=r^{\frac{m(m+1)}{2}}\prod_{l=0}^{m}\frac{1-r^{n-l}}{1-r^{l+1}}.
%\end{split}
\end{eqnarray}

At $m=0$ this equality is reduced to the condition (\ref{b1}). Respectively, at
$m\in[1,n-2]$ inserting (\ref{ca3}) into (\ref{b2}) arrives at the following
relations:
\begin{eqnarray}
%\begin{split}
 \sigma_{n}^m=\nonumber\\r^{\sum_{k=1}^{m}k}\prod_{l=0}^{m}\frac{1-r^{n-1-l}}{1-r^{l+1}}
 +r^{\sum_{k=1}^{m-1}k}\prod_{l=0}^{m-1}\frac{1-r^{n-1-l}}{1-r^{l+1}}r^{n-1}=\nonumber\\
 =r^{\sum_{k=1}^{m}k}\prod_{l=0}^{m}\frac{1-r^{n-1-l}}{1-r^{l+1}}
 \left(1+\frac{1-r^{m+1}}{1-r^{n-m-1}}r^{n-1-m}\right)\nonumber\\
 =r^{\sum_{k=1}^{m}k}\frac{\prod_{l=1}^{m+1}(1-r^{n-l})}{\prod_{l=0}^{m}(1-r^{l+1})}
 \frac{1-r^n}{1-r^{n-m-1}}\nonumber\\
 =r^{\sum_{k=1}^{m}k}\frac{\prod_{l=0}^{m}(1-r^{n-l})\frac{1-r^{n-(m+1)}}{1-r^n}}
 {\prod_{l=0}^{m}(1-r^{l+1})}\frac{1-r^n}{1-r^{n-m-1}}\nonumber\\
 =r^{\frac{m(m+1)}{2}}\prod_{l=0}^{m}\frac{1-r^{n-l}}{1-r^{l+1}}.
%\end{split}
\end{eqnarray}
In the third line, the overall multiplier is singled out off terms of the
second line; the last fraction in the fourth line is obtained with obvious
summation in brackets of the previous line; the fraction in the numerator of
the first fraction in the fifth line is appeared to single out the multipliers
related to both lower $l=0$ and upper $l=m+1$ limits in the upper production;
the last line is the result of reduction of fractions in the previous line.
Finally, at $m=n-1$ Eqs. (\ref{b3}) and (\ref{ca3}) take the equal form
\begin{equation}
 \sigma_{n}^{n-1}=r^{\frac{n(n-1)}{2}}.
 \label{cb3}
\end{equation}
Thus, one can conclude the proposition (\ref{ca3}) is applicable for all
indexes $m\in[0,n-1]$ and its insertion into Eq. (\ref{Sigma}) arrives at the
final expression of the sum of the terms of a geometrical progression
(\ref{a1}):
\begin{equation}
 S_n= a\sum\limits_{m=0}^{n-1}C_n^{m+1}(r)r^{\frac{m(m+1)}{2}}\left[(1-q)a\right]^m
 \label{Sigma2}
\end{equation}
with coefficients
\begin{equation}
 C_n^m(r)\equiv\prod_{l=0}^{m-1}\frac{1-r^{n-l}}{1-r^{l+1}}.
 \label{Sigma21}
\end{equation}

The expression (\ref{Sigma2}) can be written within the production
representation according to the relations
\begin{eqnarray} \label{S2}
%\begin{split}
S_n=\frac{1}{1-q}\left\{\sum\limits_{m=0}^{n}{C_n^{m}(r)}r^{\frac{{m(m-1)}}{2}}
[(1-q)a]^{m}-1\right\}\nonumber\\
=\frac{1}{1-q}\left\{\prod_{m=0}^{n-1}\Bigl[1+a(1-q)r^m\Bigl]-1\right\},
% \end{split}
\end{eqnarray}
the second of which expresses the deformed Gauss polynomials \cite{QG,Exton}
\begin{eqnarray}
%\begin{split}
\sum\limits_{m=0}^{n}{C_n^{m}(r)}r^{\frac{{m(m-1)}}{2}}
[(1-q)a]^{m}\nonumber\\=\prod_{m=0}^{n-1}\Bigl[1+a(1-q)r^m\Bigl].
%\end{split}
 \label{q-fact}
\end{eqnarray}
With accounting the definition (\ref{77a}), one obtains at $r=1$
\begin{equation} \label{S2-1}
S_n=\frac{{\left[{1+(1-q)a}\right]^n-1}}{{1-q}}=n\odot_qa.
\end{equation}
In non-deformed limit $q\to1$, this relation takes the trivial form $S_n=na$.

Rewriting the definition (\ref{Sigma21}) in the forms
\begin{eqnarray} \label{k5}
%\begin{split}
C_n^m(r)=\frac{{\prod\limits_{l=(n-m)+1}^n{\left({1-r^l}\right)}}}
{{\prod\limits_{l=1}^m{\left({1-r^l}\right)}}}\nonumber\\
=\frac{{\prod\limits_{l=1}^n{\left({1-r^l}\right)}}}{{\prod\limits_{l=1}^m{\left({1-r^l}\right)}
\prod\limits_{l=1}^{n-m}{\left({1-r^l}\right)}}},
%\end{split}
\end{eqnarray}
one can see that coefficients (\ref{Sigma21}) are reduced to the deformed
binomial coefficients \cite{QG,Exton}
\begin{equation}
 C_n^m(r)\equiv\frac{[n]_r!}{[m]_r![n-m]_r!}
 \label{Sigma21def}
\end{equation}
determined with the deformed factorial
\begin{eqnarray}
%\begin{split}
 [n]_r!\equiv[1]_r[2]_r\dots[n]_r\nonumber\\=\frac{(r-1)(r^2-1)\dots(r^n-1)}{(r-1)^n},
 \nonumber\\\mathstrut[n]_r\equiv1+r+r^2+\dots+r^{n-1}=\frac{r^n-1}{r-1}
%\end{split}
 \label{q-fact}
\end{eqnarray}
According to the formula (\ref{Sigma21def}), the deformed binomial coefficients
obey the usual property
\begin{equation}
C_n^m(r)=C_n^{n - m}(r). \label{k6}
\end{equation}

After replacing index $m+1$ by $m$ in Eq. (\ref{b2}) with accounting Eq.
(\ref{ca3}), we arrive at the deformed Pascal identity
\begin{equation}
C_n^m(r)=C_{n-1}^m(r)+C_{n-1}^{m-1}(r)r^{n-m}
 \label{k8}
\end{equation}
that forms the deformed Pascal triangle
\begin{eqnarray}
%\begin{split}
%\setcounter{MaxMatrixCols}{26}
\begin{array}{cccccccccc c}
%\begin{matrix}
&&&&& 1 &&&&&\\ &&&& 1 && 1 &&&&\\ &&& 1 && C_2^1(r) && 1 &&&\\ &&
1 && C_3^1(r) && C_3^1(r) && 1 &&\\ &1 && C_4^1(r) && C_4^2(r) &&
C_4^1(r) && 1&\\ 1 && C_5^1(r) && C_5^2(r) && C_5^2(r) && C_5^1(r)
&& 1
%\end{matrix}\\
\end{array}\nonumber\\
\dots
%\end{split}
\end{eqnarray}
where we put $C_n^0(r)=1$. On the other hand, iteration of the
relation (\ref{b2}) yields the sum rule
\begin{eqnarray}
%\begin{split}
C_n^m(r)=C_{n-1}^m(r)
+\left[C_{n-2}^{m-1}(r)+C_{n-2}^{m-2}(r)r^{n-m}\right]r^{n-m}\nonumber\\=\dots
=\sum\limits_{l=0}^m{C_{n-(l+1)}^{m-l}(r)r^{(n- m)l}}.
%\end{split}
 \label{k9}
\end{eqnarray}

Finally, in limit $r\to 1$ the relation (\ref{Sigma21def}) takes ordinary form:
\begin{eqnarray}
%\begin{split}
\lim_{r\to1}C_n^m(r)=\lim_{r\to1}\prod_{l=0}^m{\frac{{1-r^{n-l}}}{{1-r^{l+1}}}}\nonumber\\
=\prod_{l=0}^{m-1}{\frac{{n-l}}{{l+1}}}\equiv\frac{{n!}}{{m!(n-m)!}}.
%\end{split}
 \label{k2}
\end{eqnarray}

\end{document}